\documentclass[journal,onecolumn,12pt]{IEEEtran}

\normalsize

\usepackage{cite}
\usepackage{amsmath}
\usepackage{subcaption}
\usepackage{amssymb}
\usepackage{amsthm} 
\usepackage{epstopdf}
\usepackage{booktabs}
\usepackage[table,xcdraw]{xcolor}
\usepackage{graphicx}
\usepackage{float}
\usepackage{color}
\usepackage{latexsym}
\usepackage{algorithm}

\usepackage{caption}
\usepackage{subcaption}
\usepackage{optidef}
\usepackage[noend]{algpseudocode}
\usepackage[nodisplayskipstretch]{setspace}

\ifCLASSINFOpdf
\else
\fi

\begin{document}
	\doublespacing
	\title{Centralized \& Distributed Deep Reinforcement Learning Methods for Downlink Sum-Rate Optimization}
	
\author{Ahmad~Ali~Khan,~\IEEEmembership{Student Member,~IEEE,}
	Raviraj~Adve,~\IEEEmembership{Fellow,~IEEE,}
	\thanks{The authors are with the Department
		of Electrical and Computer Engineering, University of Toronto, Ontario,
		ON M5S 3G4, Canada. E-mails: (akhan, rsadve)@ece.utoronto.ca. This work has been accepted for publication in \textit{IEEE Transactions on Wireless Communications}.
	
		The authors would like to acknowledge the support of TELUS Canada and the National Science and Engineering Research Council, Canada through its Collaborative Research and Development (CRD) program.}% <-this % stops a space
}
	
	\maketitle
	\vspace{-6.50em}
	
	\begin{abstract}
		 \color{black}For a multi-cell, multi-user, cellular network downlink sum-rate maximization through power allocation is a nonconvex and $\mathrm{NP}$-hard optimization problem. In this paper, we present an effective approach to solving this problem through single- and multi-agent actor-critic deep reinforcement learning (DRL). Specifically, we use finite-horizon trust region optimization. Through extensive simulations, we show that we can simultaneously achieve higher spectral efficiency than state-of-the-art optimization algorithms like weighted minimum mean-squared error (WMMSE) and fractional programming (FP), while offering execution times more than two orders of magnitude faster than these approaches. Additionally, the proposed trust region methods demonstrate superior performance and convergence properties than the Advantage Actor-Critic (A2C) DRL algorithm. In contrast to prior approaches, the proposed decentralized DRL approaches allow for distributed optimization with limited CSI and controllable information exchange between BSs while offering competitive performance and reduced training times.
	\end{abstract}
	\vspace{-1.00em}
	
	\section{Introduction}
	Improving spectral efficiency is a central focus of modern research in wireless communication. As device density grows and more consumer-centric, data-intensive, applications become popular, it is clear that coordinated resource allocation is necessary to achieve this goal. Unfortunately, developing effective techniques remains fundamentally challenging due to the nature of sum-rate maximization: not only is the associated optimization problem for the broadcast channel non-convex, but, as proved by Luo and Zhang in \cite{luo2008dynamic}, it is also $\mathrm{NP}$-hard. It follows that methods to find the globally optimal solution to this problem, such as outer polyblock approximation, are computationally prohibitive and generally impractical for cellular networks involving multiple cells and multiple users \cite{liu2012achieving}.
	
	The solutions proposed in the literature balance a tradeoff between exchange of channel state information (CSI), computational complexity and performance. On the one hand, heuristic methods such as random, greedy and max-power allocation are extremely simple to implement and require no exchange of CSI between cooperating base stations \cite{sun2018learning}. However, since they ignore the impact of inter-cell and intra-cell interference they tend to have a correspondingly poor performance. 
	
	In contrast, centralized optimization algorithms like weighted minimum mean-squared error minimization (WMMSE) and fractional programming (FP) offer much better performance since they take interference into account. In particular, the work by Shi et al. in \cite{shi2011iteratively} shows the equivalence of weighted sum-rate maximization and weighted mean-squared error minimization and develops an iterative algorithm for finding a resource allocation strategy. Somewhat similarly, in \cite{shen2018fractional}, Shen and Yu develop the fractional programming approach, based on minorization-maximization methods, to solve the downlink power allocation problem iteratively. Both of these approaches are guaranteed to converge to a local optimum of the original max sum-rate problem, and as such offer excellent performance. At the same time, this performance comes with two undesirable compromises: first, these algorithms need to optimize across all cooperating BSs and thus have very high (albeit polynomial-time) computational complexity; and second, they require extensive exchange of network CSI between cooperating BSs. In other words, they assume that all the cooperating base stations are connected to a central cloud processer via high-speed backhaul links that receives the complete downlink CSI from the BSs and computes and forwards their coordinated power allocation based on this information exchange \cite{khan2018optimizing}. As a result, due to both the computational complexity and CSI exchange requirements, these methods do not scale well for deployment in real-world wireless networks, where ensuring this stringent level of cooperation and shared information across a large number of cells is infeasible. To the best of our knowledge, FP is the state-of-the art among locally optimal schemes \cite{khan2018optimizing}.
	
	{\color{black}
	To reduce the information exchange requirements, distributed optimization algorithms have been studied extensively in the literature. One approach is to model the sum-rate maximization problem as a noncooperative game, in which the BSs attempt to reach a socially acceptable power allocation strategy through best-response dynamics without having access to the complete network CSI \cite{menache2011network},\cite{de2015best}. The work in \cite{menache2011network}, for example, analyzes the conditions under which a Nash equilibrium can be reached for a discretized variant of the sum-rate maximization problem in which BSs have limited access to the network CSI. On the other hand, as noted by the authors, Nash equilibria and Pareto optimal points are not necessarily global or even local optima of the problem; thus, game-theoretic distributed optimization methods tend to have noticeably worse performance than centralized methods like FP and WMMSE. This is particularly true when BSs have access to only partial or imperfect CSI \cite{de2015best}.
	
	Distributed methods for wireless resource management have also been developed under the frameworks of robust and stochastic optimization \cite{tervo2017distributed,fritzsche2013distributed,park2010distributed}. Similar to game-theoretic approaches, however, these methods achieve performance inferior to centralized optimization approaches as the number of users becomes larger \cite{park2010distributed}. It is worth emphasizing that distributed optimization implementations of high-performance methods like FP and WMMSE are not possible: as observed in \cite{khan2018optimizing}, convergence to a local optimum of the sum-rate maximization problem is only guaranteed if the BSs have access to the full network CSI.}
	
	Resource allocation schemes based on deep supervised learning have been recently explored as a means to overcome some of these challenges \cite{sun2018learning}, \cite{d2019uplink}, \cite{matthiesen2019deep}. In \cite{sun2018learning}, Sun et al. utilized supervised regression to enable a deep feedforward neural network to approximate the locally optimal power allocation solution achieved by the WMMSE algorithm. Similarly, in \cite{d2019uplink}, the authors utilize a deep supervised learning algorithm to approximate the \textit{globally} optimal power allocation strategy for sum-rate maximization in the uplink of a massive MIMO network. In both of these cases, the nonlinear function approximation property of neural networks was successfully leveraged to learn a mapping from the input CSI to the desired solution of the problem. 
	
	While such an approach reduces execution time once the trained model is deployed, it also suffers from an obvious drawback: the level of CSI exchange required is identical to the optimization algorithms supervised learning aims to mimic. A further disadvantage is that the supervised learning algorithm in \cite{sun2018learning} can \textit{at best} match the performance of the WMMSE algorithm, not exceed it. For the approach used in \cite{d2019uplink}, a major concern is generating samples of the globally optimal solution for the algorithm to learn from, since, as mentioned earlier, the original optimization problem is $\mathrm{NP}$-hard. In \cite{matthiesen2019deep}, Mattheisen et al. aim to mitigate this problem by using branch-and-bound to find globally optimal solutions to the (also $\mathrm{NP}$-hard) energy efficiency maximization problem; however, the computational cost of generating a sufficiently large number of samples to enable effective supervised learning for a deep neural network remains extremely high.
	
	Due to the issues associated with supervised learning methods, deep reinforcement learning (DRL) has emerged as a viable alternative to solve wireless resource management problems \cite{nasir2018deep,Meng2019,wei2017user,zhang2019calibrated}. Crucially, DRL methods do not require prior solutions to the problem unlike supervised learning methods; instead, the algorithms improve over time through a mechanism of feedback and interaction. For example, in \cite{nasir2018deep}, the authors present a multi-agent deep Q-learning (DQL) approach to solve the problem of link spectral efficiency maximization; the proposed approach provides performance that closely matches the FP algorithm but requires discretization of the power control variables. In a similar fashion, the work in \cite{Meng2019} explores the use of DQL, REINFORCE, and deep deterministic policy gradient (DDPG) algorithms for link sum-rate maximization. The DQL and REINFORCE algorithms presented once again require discretization of the power allocation variables; this introduces uncertainty since there is no known method for choosing the optimal discretization factor.
	
	 {\color{black}A major issue with the approaches presented in \cite{nasir2018deep} and \cite{Meng2019} is that direct optimization of the network spectral efficiency is not possible; instead, as a proxy, each transmitter aims to maximize the difference between its own spectral efficiency and the weighted leakage to other transmitters}. This proxy objective function is only proportional to the sum-rate when the number of transmitter-receiver pairs is very large. Additionally, the approaches proposed require CSI exchange between the transmitters to function. The learning algorithms presented also require significant feature engineering; for example, in \cite{Meng2019}, in addition to the current CSI, historical power allocation values are needed in order to output current power allocation values; this adds additional requirements to the already burdensome information exchange problem.
	 
	 {\color{black}
	 Finally, we observe that the aforementioned deep learning methods do not allow for distributed optimization with limited CSI available at each BS. For example, the supervised learning method applied in \cite{sun2018learning} to approximate the WMMSE algorithm can only be implemented in a centralized fashion which, once again, requires complete CSI exchange between all BSs in the network. As the authors of \cite{sun2018learning} demonstrate, supervised learning would be unable to succeed in approximating the WMMSE algorithm in a decentralized environment, since there would not be a one-to-one mapping between the CSI input and power allocation output at each BS. Similarly, the reinforcement learning approaches implemented in \cite{nasir2018deep} and \cite{Meng2019} require full CSI exchange between BSs, in addition to prior channel state information (and per-cell spectral efficiency) from previous time slots. Furthermore, to compute the power allocation for each link, the transmitter requires a fixed number of the strongest interference channels as input; this necessitates the exchange of CSI between BSs even after the agents have been deployed and training is complete. These processing and information exchange requirements necessarily prevent the implementation of distributed optimization of the network spectral efficiency. Thus, we conclude that the current optimization approaches, both learning-based and otherwise, compromise by either having prohibitively high computation and information exchange requirements or poor performance. It is this gap in the literature that we aim to address with our work.}
	
	In this paper, we present a class of DRL methods based on Markov Decision Processes (MDP), \textit{trust-region} policy optimization (TRPO) for downlink power allocation. In contrast to the deep Q-learning method, trust region methods are capable of directly solving continuous-valued problems; this removes the problem of decision spaces growing exponentially larger as the discretization of the optimization variables is increased. Compared to the REINFORCE approach, trust region methods demonstrate greater robustness since we can ensure that updates meet practically justifiable constraints; thus, they have been shown to achieve far higher performance across a variety of continuous-control reinforcement learning problems \cite{achiam2017constrained}, \cite{schulman2015trust}. {\color{black}Specifically, we utilize finite-horizon trust region methods to derive centralized and decentralized algorithms for sum-rate optimization which require varying degrees of CSI exchange between BSs.}
	
	The contributions of this paper can be summarized as follows:
	\begin{itemize}
		\item We demonstrate that the use of finite-horizon trust region policy optimization is effective in directly solving the sum-rate maximization problem in a multi-cell, multi-user environment. Unlike previous deep RL-based attempts which necessitate discretization of the power variables and the use of proxy objective functions, we solve the \textit{continuous-valued} power allocation problem with the network sum-rate taken directly as the reward function, thereby simplifying the algorithm design and implementation.
		\item The approaches we present provide notably higher spectral efficiency than the state-of-the-art FP and WMMSE algorithms across different network sizes.
		\item We show that agents trained using the proposed approach produce solutions to the sum-rate maximization problem over two orders of magnitude faster than the aforementioned conventional optimization algorithms. {\color{black}  The proposed TRPO methods also demonstrate lower variance and higher performance than the conventional Advantage Actor-Critic (A2C) algorithm.}
		\item {\color{black}The multi-agent distributed approaches allow us to flexibly control information exchange between BSs, which is not possible in prior distributed optimization methods or using prior deep learning approaches. In particular, the partially decentralized strategy we propose allows BSs to exchange only power allocation information to effectively improve network sum-rate while utilizing only \textit{local} CSI at each BS.
		\item The trained DRL methods demonstrate robust performance to changes in the reward function (i.e., the network sum-rate) across a range of BS transmit powers.}
	\end{itemize}
	
	This paper is organized as follows: in Section II, we present the system model and formulate the network sum-rate maximization problem. In Section III, we describe the Markov Decision Process (MDP) framework, and derive the proposed centralized and decentralized DRL algorithms. This is followed by training, performance, and execution time results and comparisons in Section IV. We draw some conclusions in Section V.
	\section{System Model and Problem Formulation}
	Our system model is similar to that used in \cite{shen2018fractional}. We consider a time-division duplexed network comprising of $B$ single-antenna base stations and $K$ single-antenna users per cell uniformly distributed over the cell area. {\color{black}The combined downlink channel gain from BS $b$ to user $k$ associated with base station $b'$ in the $n^{th}$ time slot  is denoted by ${h}_{{b}\rightarrow{k}{,}{b}{'}}^{(n)}$ and given by}
	\[{h}_{{b}\rightarrow{k}{,}{b}{'}}^{(n)}{=}{g}_{{b}\rightarrow{k}{,}{b}{'}}^{(n)}\sqrt{{\mathit{\beta}}_{{b}\rightarrow{k}{,}{b}{'}}^{(n)}}\]
	where ${g}_{{b}\rightarrow{k}{,}{b}{'}}^{(n)}\sim{\mathcal{CN}}{(}{0}{,}{1}{)}$ represents the complex-valued small-scale Rayleigh fading component and ${\mathit{\beta}}_{{b}\rightarrow{k}{,}{b}{'}}^{(n)}$ represents the real-valued pathloss component. The latter is given by:
	\[{\mathit{\beta}}_{{b}\rightarrow{k}{,}{b}{'}}^{(n)}{=}{\left({{1}{+}\frac{{d}_{{b}\rightarrow{k}{,}{b}{'}}^{(n)}}{{d}_{0}}}\right)}^{{-}\mathit{\alpha}}\]
	where ${d}_{{b}\rightarrow{k}{,}{b}{'}}^{(n)}$ denotes the Euclidean distance in meters between the base station and user in question, $d_0$ is a reference distance and $\alpha$ is the pathloss exponent. Hence, ${h}_{{b}\rightarrow{k}{,}{b}}^{(n)}$ denotes the information-bearing channel from BS $b$ to user $k$ within its cell while ${h}_{{b}\rightarrow{k}{,}{b}{'}}^{(n)}$ indicates the interference channel from the same BS to user $k$ being served by BS $b'$. {\color{black}Furthermore, we assume that the Rayleigh fading coefficients of the information-bearing and interference channels are independent across users and time slots.}
	
	{\color{black}
	 For notational convenience, the network CSI in the $n^{th}$ time slot is denoted by the $K\times{B}\times{B}$ tensor ${\mathbf{H}}^{(n)}$, i.e.,
	\[
	{\mathbf{H}}^{(n)}{=}\left\{{{h}_{{b}\rightarrow{k}{,}{b}{'}}^{(n)}{|}{b}{,}{b}{'}{=}{1}{,}\ldots{,}{B}{;}\hspace{0.33em}{k}{=}{1}{,}\ldots{,}{K}}\right\}
	\]
	The transmit power allocated by BS $b$ to user $k$ within its cell is denoted by $p_{b\rightarrow{k}}$ and is similarly collected in the $K\times{B}$ matrix ${\mathbf{P}}^{(n)}$.} Assuming a receiver noise power of ${z}_{k,b}^{(n)}$ at user $k$ served by BS $b$, the network sum-rate for the $n^{th}$ time slot is given by:
	{\color{black}
	\[
{R}\left({{\mathbf{H}}^{(n)},{\mathbf{P}}^{(n)}}\right){=}\mathop{\sum}\limits _{\left({b,k}\right)}{{\log}_{2}\left({{1}{+}\frac{{p}_{{b}\rightarrow{k}}^{(n)}{\left|{{h}_{{b}\rightarrow{k}{,}{b}}^{(n)}}\right|}^{2}}{\sum_{(b',k')\neq(b,k)}{p}_{{b}{'}\rightarrow{k}}^{(n)}{\left|{{h}_{{b}{'}\rightarrow{k}{,}{b}}^{(n)}}\right|}^{2}{+}{z}_{k,b}^{(n)}}}\right)}
	\]
}
	Based on these definitions, the goal of maximizing the network spectral efficiency for a single time slot can be expressed as the following optimization problem:
	\begin{subequations}\label{SR_max_problem}
		\begin{align}
		{\mathop{\mathrm{maximize}}\limits_{{{\mathbf{P}}^{(n)}}}\hspace{0.33em}{R}\left({{\mathbf{H}}^{(n)},{\mathbf{P}}^{(n)}}\right)}\hspace{14.63em}\label{network_sumrate}\\
		{\mathrm{subject}\hspace{0.33em}\mathrm{to}\hspace{0.67em}{0}\leq{p}_{{b}\rightarrow{k}}^{(n)}\leq{P}_{\max}\hspace{1.33em}{k}{=}{1}{,}\ldots{,}{K}{;}\hspace{0.33em}{b}{=}{1}{,}\ldots{,}{B}}\label{power_constraint}
		\hspace{1.95em}
		\end{align}
	\end{subequations}
	where the constraint in (\ref{power_constraint}) ensures that the power allocated to each of the users is nonnegative and does not exceed the maximum allowed transmit power.
	As stated earlier, this optimization problem has been shown to be non-convex and NP-hard in~\cite{luo2008dynamic}.
	
	\section{Proposed Deep Reinforcement Learning Approach}
	\subsection{Markov Decision Process (MDP)}
	Our proposed DRL-based algorithms employ trust-region methods to solve problem (\ref{SR_max_problem}) in both single-agent centralized and multi-agent distributed fashion. Both these variants are modeled using the Markov Decision Process (MDP) framework; accordingly, in this section we define some related terminology that will be used throughout this paper.
	
\begin{figure}[!t] 
		\begin{center} 
			\includegraphics[trim={0cm 3cm 0cm 3cm},clip, height=0.30\textwidth]{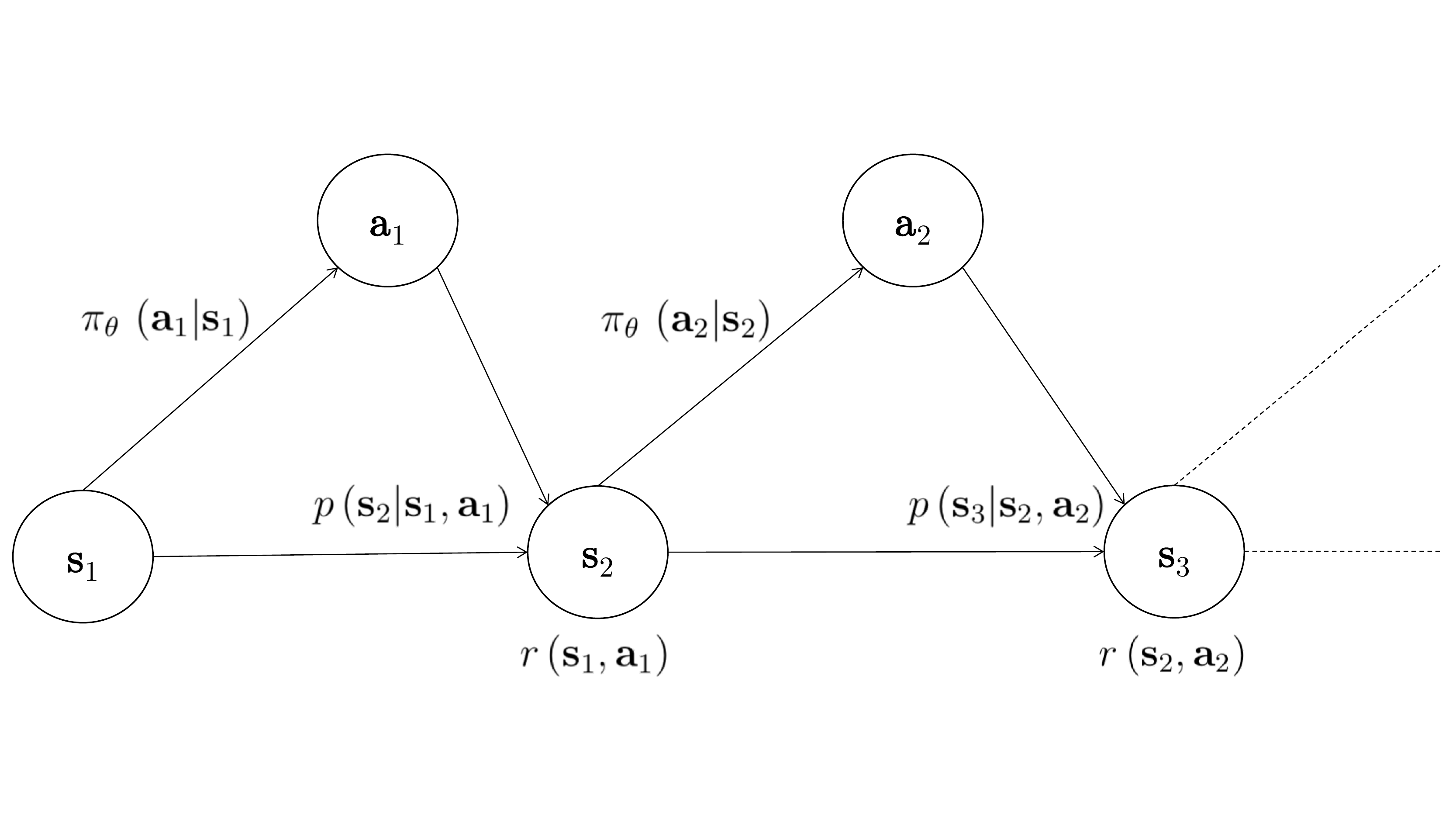}		
			\caption{Graphical model illustrating transitions in a Markov Decision Process.}
			\centering
			\label{MDP}
		\end{center}
	\end{figure}

	An \textit{agent} is an entity capable of processing information from its environment and taking decisions directed towards the maximization of a chosen reward function. The agent interacts with its environment at discrete time instants ${n}{=}{1}{,}\ldots{,}{N}$ (where $N$, the \textit{episode length}, is fixed). The interaction model of the agent with its enivronment can be described as follows:
	
	\begin{itemize}
		\item At time step $n$, the agent is assumed to be in \textit{state} ${\mathbf{s}}_{n}\in{\mathcal{S}}$. The state comprises the information that the agent has access to and defines its situation relative to its environment. The \textit{state space}, $\mathcal{S}$, is defined as the set of all possible states that can be encountered by the agent.
		\item Acting only upon the state at time step $n$, the agent takes \textit{action} ${\mathbf{a}}_{n}\in{\mathcal{A}}$ from the set of all possible actions i.e., the \textit{action space ${\mathcal{A}}$}. As illustrated in Figure \ref{MDP}, the action is sampled from a conditional probability distribution known as the \textit{policy}, denoted by ${\mathit{\pi}}_{\boldsymbol{\theta}}\left({{\mathbf{a}}_{n}|{\mathbf{s}}_{n}}\right)$, where the subscript indicates that the policy is parameterized through the (generally tensor-valued) variable $\boldsymbol{\theta}$. In this work, we use the terms \textit{policy} and \textit{policy network} interchangeably since a deep neural network is used to parametrize the policy.
		\item We assume a Markov state transition model; thus, the probability of transition to ${\mathbf{s}}_{n+1}$ is assumed to depend only on the current state, ${\mathbf{s}}_{n}$, and current action, ${\mathbf{a}}_{n}$:
		\begin{equation}\label{Markov_transition}
		{p}\left({{\mathbf{s}}_{{n}{+}{1}}{|}{\mathbf{s}}_{n,}{\mathbf{a}}_{n}{,}\ldots{,}{\mathbf{s}}_{1}{,}{\mathbf{a}}_{1}}\right){=}{p}\left({{\mathbf{s}}_{{n}{+}{1}}|{\mathbf{s}}_{n,}{\mathbf{a}}_{n}}\right) 
		\end{equation}
		{\color{black}Note that the transition probability can be completely deterministic based on the agent's current action; likewise, it may be completely independent of the current action \cite{arulkumaran2017deep}}. This Markov transition property is depicted further in Figure \ref{MDP}. The state transition likelihoods are contained in the transition operator $\mathcal{T}$. Furthermore, the state distribution arising as a result of the policy ${\mathit{\pi}}_{\boldsymbol{\theta}}$ and environmental transition dynamics is denoted as ${d}_{{\mathit{\pi}}_{\boldsymbol{\theta}}}$ and referred to as the \textit{policy state distribution}.	
		\item At the end of the time step, the agent receives a scalar \textit{reward}, ${r}{(}{\mathbf{s}}_{n}{,}{\mathbf{a}}_{n})$, which is a function of the current state and current action, i.e.,  ${r}{(}{\mathbf{s}}_{n}{,}{\mathbf{a}}_{n}{):}\hspace{0.33em}{\mathcal{S}}\times{\mathcal{A}}\mapsto{\mathbb{R}}$.
	\end{itemize}
	Taken together, the tuple ${\mathcal{M}}{=}\left\{{\mathcal{S},\mathcal{A},\mathcal{T},r}\right\}$ defines an MDP. Using the Markov property of state transitions, the likelihood of observing a given sequence of states and actions in the MDP can be decomposed as:
	\begin{equation}
	{p}\left({{\mathbf{s}}_{1}{,}{\mathbf{a}}_{1}{,}\ldots{,}{\mathbf{s}}_{N}{,}{\mathbf{a}}_{N}}\right){=}{p}{(}{\mathbf{s}}_{1}{)}\mathop{\prod}\limits_{{n}{'}{=}{1}}\limits^{N}{{p}\left({{\mathbf{s}}_{{n}{'}{+}{1}}|{\mathbf{s}}_{n'},{\mathbf{a}}_{n'}}\right){\mathit{\pi}}_{\boldsymbol{\theta}}\left({{\mathbf{a}}_{n'}|{\mathbf{s}}_{n'}}\right)}
	\end{equation}
	
	The \textit{average discounted reward} achieved over an episode is a function of $\boldsymbol{\theta}$ and is given by 
	\begin{equation}\label{expected_reward}
	\bar{R}{(}\boldsymbol{\theta}{)}{=}\mathop{\mathbb{E}}\limits_{{\mathbf{a}}_{n'}\sim{\mathit{\pi}}_{\boldsymbol{\theta}}{,}\hspace{0.33em}{\mathbf{s}}_{n'}\sim{d}_{{\mathit{\pi}}_{\boldsymbol{\theta}}}}\left[{\mathop{\sum}\limits_{{n}{'}{=}{1}}\limits^{N}{{\mathit{\gamma}}^{n'-1}{r}\left({{\mathbf{s}}_{n'},{\mathbf{a}}_{n'}}\right)}}\right]
	\end{equation}
	where $\mathit{\gamma}\in\left[{0,1}\right]$ is a discount factor controlling the value of future rewards relative to current rewards. The subscripts ${\mathbf{a}}_{n}\sim{\mathit{\pi}}_{\boldsymbol{\theta}}$ and $
	{\mathbf{s}}_{n}\sim{d}_{{\mathit{\pi}}_{\boldsymbol{\theta}}}
	$ indicate that the expectation is with respect to actions sampled from the policy and states sampled from the policy state distribution.
	
	We also define the \textit{state-action value function}, the \textit{state value function} and the \textit{advantage function}. The state-action value function ${Q}^{{\mathit{\pi}}_{\boldsymbol{\theta}}}\left({{\mathbf{s}}_{n},{\mathbf{a}}_{n}}\right)
	$ (also known as the $Q$-function), is the total expected reward that can be accumulated by taking action ${\mathbf{a}}_{n}$ from state ${\mathbf{s}}_{n}$ and following the policy ${\mathit{\pi}}_{\boldsymbol{\theta}}$ in subsequent timesteps, i.e.,
	\begin{equation}\label{state_action_value}
	{Q}^{{\mathit{\pi}}_{\boldsymbol{\theta}}}\left({{\mathbf{s}}_{n},{\mathbf{a}}_{n}}\right){\triangleq}\mathop{\mathbb{E}}\limits_{{\mathbf{a}}_{n'}\sim{\mathit{\pi}}_{\boldsymbol{\theta}}{,}\hspace{0.33em}{\mathbf{s}}_{n'}\sim{d}_{{\mathit{\pi}}_{\boldsymbol{\theta}}}}\left[{\mathop{\sum}\limits_{{n}{'}{=}{n}}\limits^{N}{{\mathit{\gamma}}^{{n}{'}{-}{n}}{r}\left({{\mathbf{s}}_{n'},{\mathbf{a}}_{n'}}\right){|}}{\mathbf{s}}_{n},{\mathbf{a}}_{n}}\right]
	\end{equation}
	
	The expectation of the state-action value function, ${V}^{{\mathit{\pi}}_{\boldsymbol{\theta}}}\left({{\mathbf{s}}_{n}}\right)$, over all possible actions with respect to the policy and its corresponding state distribution is defined as the state-value function and is given by:
	\begin{equation}\label{state_value}
	{V}^{{\mathit{\pi}}_{\boldsymbol{\theta}}}\left({{\mathbf{s}}_{n}}\right){\triangleq}\mathop{\mathbb{E}}\limits_{{\mathbf{a}}_{n'}\sim{\mathit{\pi}}_{\boldsymbol{\theta}}{,}\hspace{0.33em}{\mathbf{s}}_{n'}\sim{d}_{{\mathit{\pi}}_{\boldsymbol{\theta}}}}\left[{\mathop{\sum}\limits_{{n}{'}{=}{n}}\limits^{N}{{\mathit{\gamma}}^{{n}{'}{-}{n}}{r}\left({{\mathbf{s}}_{n'},{\mathbf{a}}_{n'}}\right){|}{\mathbf{s}}_{n}}}\right]{=}\mathop{\mathbb{E}}\limits_{{\mathbf{a}}_{n}\sim{\pi}_{\boldsymbol{\theta}}}{\left[{{Q}^{{\pi}_{\boldsymbol{\theta}}}\left({{\mathbf{s}}_{n},{\mathbf{a}}_{n}}\right)}\right]}
	\end{equation}
	
	In other words, ${Q}^{{\mathit{\pi}}_{\boldsymbol{\theta}}}\left({{\mathbf{s}}_{n},{\mathbf{a}}_{n}}\right)
	$ indicates the expected value of choosing a particular action ${\mathbf{a}}_{n}$ in time step $n$ and afterwards following the policy whereas ${V}^{{\mathit{\pi}}_{\boldsymbol{\theta}}}\left({{\mathbf{s}}_{n}}\right)$ indicates the expected reward that can be obtained by simply following the policy throughout.
	
	 The difference between the state-action value function and state-value function is called the advantage function and is denoted by ${A}^{{\mathit{\pi}}_{\boldsymbol{\theta}}}\left({{\mathbf{s}}_{n},{\mathbf{a}}_{n}}\right)$:
	\begin{equation}\label{advantage}
	{A}^{{\mathit{\pi}}_{\boldsymbol{\theta}}}\left({{\mathbf{s}}_{n},{\mathbf{a}}_{n}}\right){\triangleq}{Q}^{{\mathit{\pi}}_{\boldsymbol{\theta}}}\left({{\mathbf{s}}_{n},{\mathbf{a}}_{n}}\right){-}{V}^{{\mathit{\pi}}_{\boldsymbol{\theta}}}\left({{\mathbf{s}}_{n}}\right)
	\end{equation}
	
	Intuitively, the advantage function can be interpreted as a measure of how much better an action ${\mathbf{a}}_{n}$ is than the average action chosen by the policy ${\mathit{\pi}}_{\boldsymbol{\theta}}$.
	
	The goal in reinforcement learning is to find the optimum value of the policy variable $\boldsymbol{\theta}$ that leads to maximization of the expected reward:
	\begin{equation} \label{policy_optimization_problem}
	{\boldsymbol{\theta}}^{\mathbf{*}}{=}\mathop{\arg\max}\limits_{\hspace{2.33em}\boldsymbol{\theta}{\mathbf{'}}}\hspace{0.33em}\bar{R}{(}\boldsymbol{\theta}{\mathbf{'}}{)}
	\end{equation}
	
	Direct differentiation of the expected reward $\bar{R}{(}\boldsymbol{\theta}{)}$ with respect to the policy parameter yields a closed-form expression for the \textit{policy gradient}, a Monte-Carlo sampled estimate of which can then be used to update the policy through the well-known REINFORCE algorithm as developed by Williams \cite{williams1992simple}. In other words, the REINFORCE algorithm attempts to change the policy parameter in the direction of increasing reward as obtained from samples of the policy. Actor-critic algorithms employ a similar approach, but utilize a separate function approximator (typically also a deep neural network with parameters $\boldsymbol{\phi}$), called the \textit{value network} to estimate the state-action value function ${Q}^{{\mathit{\pi}}_{\boldsymbol{\theta}}}\left({{\mathbf{s}}_{n},{\mathbf{a}}_{n}}\right)$ for gradient computation; it can be shown that this results in estimates of the gradient with lower variance and hence improved performance \cite{mnih2016asynchronous}.
	
	Unfortunately, both the REINFORCE and standard actor-critic algorithms suffer from a serious defect: choosing a suitable step size for the policy parameter update is extremely challenging. Depending on the parameterization, even small changes in the policy parameter $\boldsymbol{\theta}$ can lead to drastic changes in the policy output ${\mathit{\pi}}_{\boldsymbol{\theta}}$; this is especially true for policies utilizing deep neural networks, in which the output is a highly nonlinear function of the weights and biases. It is therefore desirable to seek algorithms that modify both the ascent direction and step size to achieve stable, incremental changes in the policy space rather than parameter space.
	
	\subsection{Centralized Trust Region Policy Optimization}
\begin{figure}[t!]
    	\centering
    	\begin{subfigure}[]{0.8\textwidth}
    		\centering
    		\includegraphics[trim={0cm 0cm 0cm 0cm},clip, height=0.5\textwidth]{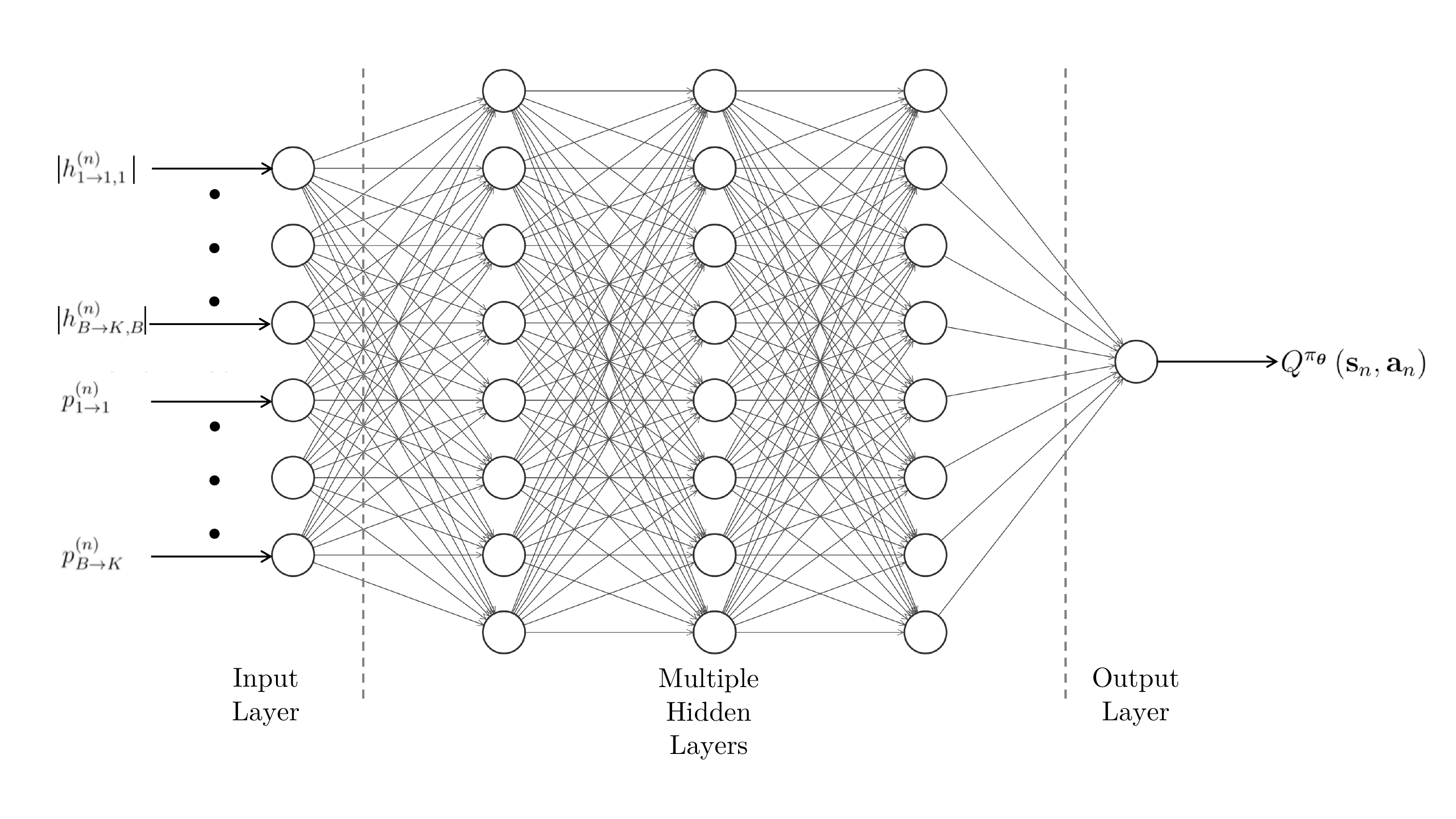}
    		\caption{Value Network.}
    	\end{subfigure}
    \begin{subfigure}[]{0.8\textwidth}
    		\centering
    		\includegraphics[trim={0cm 0cm 0cm 0cm},clip, height=0.5\textwidth]{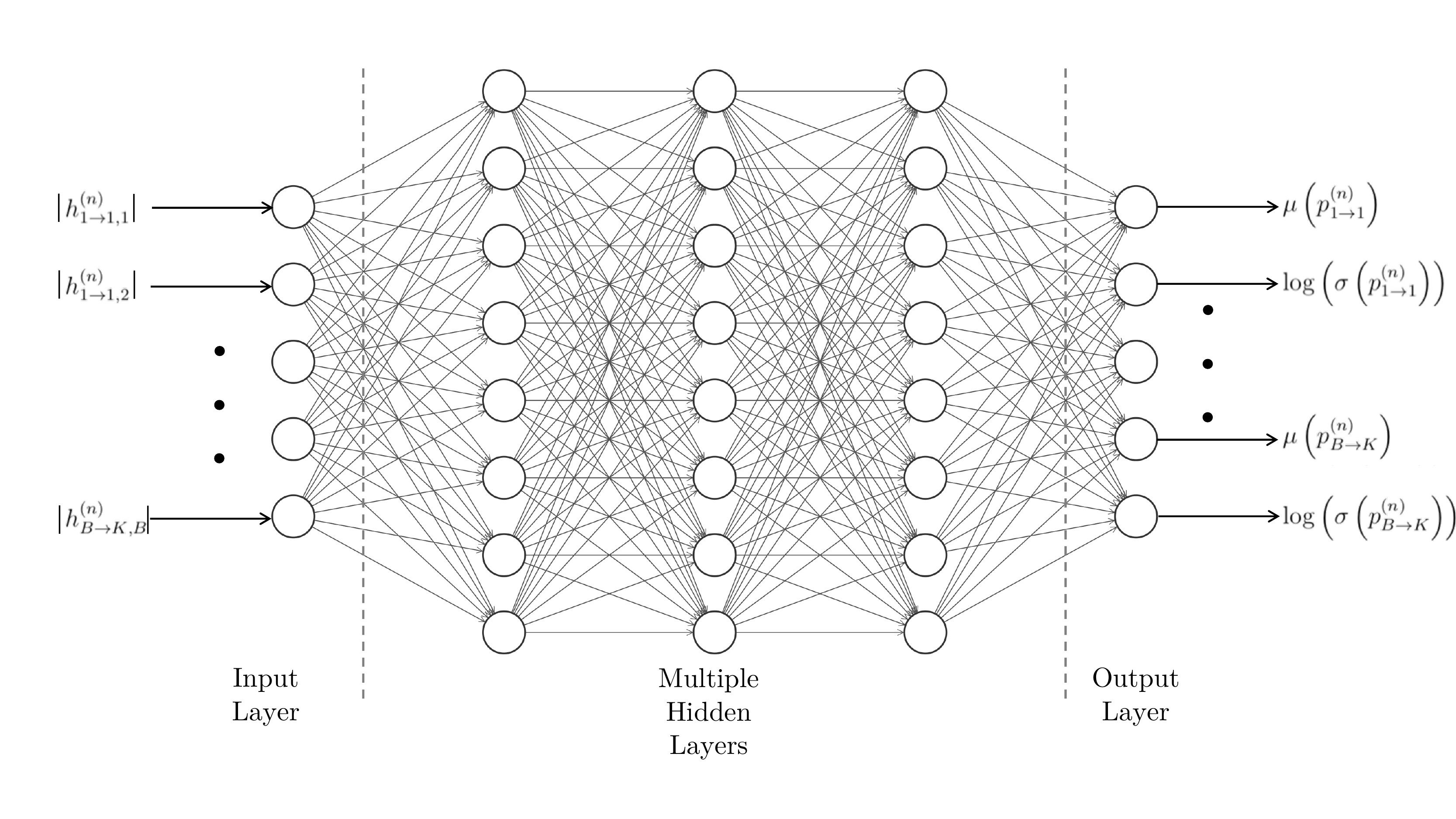}
    		\caption{Policy Network.}
    	\end{subfigure}
    	\caption{The value network outputs an estimate of the value function; in contrast, the policy network produces a mean and log-standard deviation of the action as output.}\label{value_policy_networks}
    \end{figure}

	In this section, we consider the application of trust-region policy optimization to the sum-rate maximization problem. We begin by designing a centralized approach in which a single agent (i.e., a policy network with parameter $\boldsymbol{\theta}$) receives the downlink CSI from all base stations, computes the network power allocation strategy and forwards it to the base stations. 
	\begin{figure}[!t] 
		\begin{center} 
			\includegraphics[trim={8cm 3cm 12cm 3cm},clip, height=0.31\textwidth]{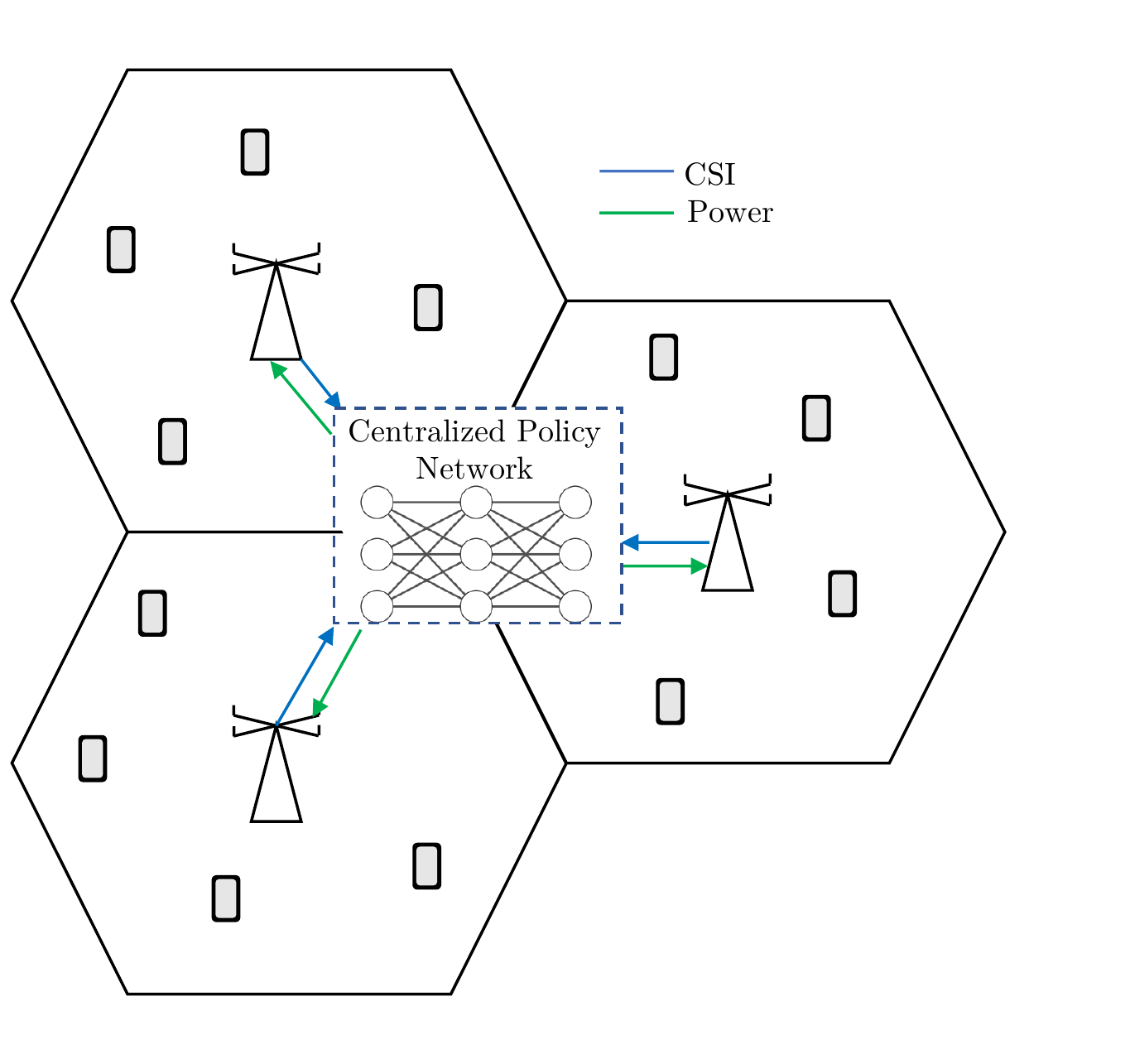}		
			\caption{For the centralized approach, each BS forwards its downlink CSI to a policy network which then determines the power allocation strategy for the entire network.}
			\centering
			\centering
			\label{centralized_diagram}
		\end{center}
	\end{figure} 
	Thus, it follows that the state and action for the centralized agent are the complete network CSI and power allocation matrix respectively, i.e., 	${\mathbf{s}}_{n}{=}{\mathbf{H}}^{{\left({n}\right)}}$ and ${\mathbf{a}}_{n}{=}{\mathbf{P}}^{{\left({n}\right)}}$.
    We parameterize a probabilistic policy over the actions as follows: the policy network takes as input the current state and outputs the mean and log-standard deviation values for each user's normally-distributed power allocation, as illustrated in Figure \ref{value_policy_networks}. The power allocation variables are then sampled using the following normal distribution:
    \begin{equation}\label{power_sampling}
    {p}_{{b}\rightarrow{k}}^{\left({n}\right)}\sim{\left[{{\mathcal{N}}\left({\mathit{\mu}\left({{p}_{{b}\rightarrow{k}}^{\left({n}\right)}}\right){,}{\mathit{\sigma}}^{2}\left({{p}_{{b}\rightarrow{k}}^{\left({n}\right)}}\right)}\right)}\right]}_{0}^{{P}_{\max}}
    \end{equation}
    where ${\mathit{\mu}\left({{p}_{{b}\rightarrow{k}}^{\left({n}\right)}}\right)}$ and ${\mathit{\sigma}}^{2}\left({{p}_{{b}\rightarrow{k}}^{\left({n}\right)}}\right)$ are the policy network output mean and variance values for the power allocated by the $b^{th}$ BS to the $k^{th}$ user, and the limits are set to ensure that the power constraints in (\ref{power_constraint}) are enforced. Thus, the input size for the policy network is $KB^{2}$ and the output size is $2KB$. On the other hand, the value network takes the current state and action as inputs (and hence has input size $KB^{2}+2KB$) and outputs an estimate of the state-action value function (which is a scalar, hence the output size is 1).

	Our goal is to optimize the policy network parameter $\boldsymbol{\theta}$ to maximize the expected reward, which we choose as the network sum-spectral efficiency averaged across multiple time slots:
\[\bar{R}\left({\boldsymbol{\theta}}\right){=}\mathop{\mathbb{E}}\limits_{{\mathbf{a}}_{n}\sim{\pi}_{\boldsymbol{\theta}}{,}\hspace{0.33em}{\mathbf{s}}_{n}\sim{d}_{{\pi}_{\boldsymbol{\theta}}}}\left[{\frac{1}{N}\mathop{\sum}\limits_{{n}{=}{1}}\limits^{N}{{\mathit{\gamma}}^{n}{R}\left({{\mathbf{H}}^{(n)},{\mathbf{P}}^{(n)}}\right)}}\right]\]
	{\color{black}
	It should be noted that for this centralized approach, the likelihood of transition to a new state (i.e., ${\mathbf{H}}^{{\left({n}\right)}}$) in the $n^{th}$ time slot is, in fact, independent of the actions taken in the previous time slot (i.e., ${\mathbf{P}}^{{\left({n-1}\right)}}$). However, as we shall see, this does not affect the derivations of the algorithms that follow. 
	
	{\color{black}
	As illustrated in Figure \ref{centralized_diagram}, the centralized method is similar to the FP and WMMSE algorithms in terms of the computation and CSI exchange. Further, we emphasize that since we consider an episode to consist of a single time-slot, the scenario is conceptually similar to the well-known `multi-armed bandit' problem in reinforcement learning \cite{kuleshov2014algorithms} in which the episode length consists of a single time slot. This single time-slot episode approach has also been utilized in prior works exploring the use of policy-based deep reinforcement learning methods to solve resource allocation problems, such as \cite{liang2019towards} and \cite{eisen2019learning}. 
}

	To optimize the policy network parameter, we adopt trust-region methods; these alter the ascent direction while utilizing decaying step sizes to avoid destructively large policy updates \cite{schulman2015high}. By bounding the change in expected reward as a function of the change in policy parameters, we can develop an iterative optimization approach that uses second-order curvature information, rather than just the gradient, to achieve robust policy updates. Our approach can be viewed as a hybrid of the approaches presented in \cite{schulman2015trust,achiam2017constrained}: we utilize a finite-horizon approach derived from that in \cite{schulman2015trust} with the bounds established in \cite{achiam2017constrained}.}
	
	 We begin by casting the policy optimization problem in the following equivalent form: suppose that the current policy parameter is $\boldsymbol{\theta}$; then the problem of maximizing expected reward is equivalent to finding the new policy parameter $\boldsymbol{\theta}^{*}$ that maximizes the difference in expected reward over the current policy:
	\begin{equation}
	\label{difference_optimization_problem}
	{\boldsymbol{\theta}}^{\mathbf{*}}{=}\mathop{\arg.\max}\limits_{\hspace{2.39em}\boldsymbol{\theta}{\mathbf{'}}}\hspace{0.33em}\bar{R}{(}\boldsymbol{\theta}{\mathbf{'}}{)}{-}\bar{R}{(}\boldsymbol{\theta}{)}
	\end{equation}

	{\color{black}
	Instead of attempting to optimize the difference in expected reward directly, we derive the following lower bound similar to the approach in \cite{schulman2015trust}:
	\begin{subequations}
		\begin{align}
		\hspace{-2.00em}\bar{R}{(}\boldsymbol{\theta}{')}{-}\bar{R}{(}\boldsymbol{\theta}{)}{=}\mathop{\mathbb{E}}\limits_{{{{\mathbf{a}}_{n'}\sim{\mathit{\pi}}_{\boldsymbol{\theta}{\mathbf{'}}}}\atop{{\mathbf{s}}_{n'}\sim{d}_{{\mathit{\pi}}_{\boldsymbol{\theta}{\mathbf{'}}}}}}}\left[{\mathop{\sum}\limits_{{n}{'}{=}{1}}\limits^{N}{{\mathit{\gamma}}^{n'}{A}^{{\mathit{\pi}}_{\boldsymbol{\theta}}}({\mathbf{s}}_{n'},{\mathbf{a}}_{n'})}}\right]\hspace{2.60em}\label{bound1}\\
		{=}\mathop{\mathbb{E}}\limits_{{{{\mathbf{a}}_{n'}\sim{\mathit{\pi}}_{\boldsymbol{\theta}}}\atop{{\mathbf{s}}_{n'}\sim{d}_{{\mathit{\pi}}_{\boldsymbol{\theta}{\mathbf{'}}}}}}}\left[{\mathop{\sum}\limits_{{n}{'}{=}{1}}\limits^{N}{{\mathit{\gamma}}^{n'}\frac{{\mathit{\pi}}_{\boldsymbol{\theta}{'}}({\mathbf{a}}_{n'}|{\mathbf{s}}_{n'})}{{\mathit{\pi}}_{\boldsymbol{\theta}}({\mathbf{a}}_{n'}|{\mathbf{s}}_{n'})}{A}^{{\mathit{\pi}}_{\boldsymbol{\theta}}}({\mathbf{s}}_{n'},{\mathbf{a}}_{n'})}}\right]\label{bound2}\hspace{-2.10em}\\
		{\geq}{\mathcal{L}}_{\boldsymbol{\theta}}\left({\boldsymbol{\theta}{\mathbf{'}}}\right){-}{C}\sqrt{\mathop{\mathbb{E}}\limits_{{{\mathbf{s}_{n'}\sim{d}_{{\mathit{\pi}}_{\boldsymbol{\theta}}}}\atop{}}}\left[{{D}_{KL}\left({{\mathit{\pi}}_{\boldsymbol{\theta}{'}}\parallel{\mathit{\pi}}_{\boldsymbol{\theta}}}\right)}\right]}\label{bound3}
		\end{align}
	\end{subequations}
	where
	\begin{equation}\label{L}
	{\mathcal{L}}_{\boldsymbol{\theta}}\left({\boldsymbol{\theta}{\mathbf{'}}}\right)\triangleq\mathop{\mathbb{E}}\limits_{{{{\mathbf{a}}_{n'}\sim{\mathit{\pi}}_{\boldsymbol{\theta}}}\atop{{\mathbf{s}}_{n'}\sim{d}_{{\mathit{\pi}}_{\boldsymbol{\theta}}}}}}\left[{\mathop{\sum}\limits_{{n}{'}{=}{1}}\limits^{N}{{\mathit{\gamma}}^{n'}\frac{{\mathit{\pi}}_{\boldsymbol{\theta}{'}}({\mathbf{a}}_{n'}|{\mathbf{s}}_{n'})}{{\mathit{\pi}}_{\boldsymbol{\theta}}({\mathbf{a}}_{n'}|{\mathbf{s}}_{n'})}{A}^{{\mathit{\pi}}_{\boldsymbol{\theta}}}({\mathbf{s}}_{n'},{\mathbf{a}}_{n'})}}\right]
	\end{equation}
	 and the optimal value of the constant $C$ to make this inequality tight can be determined analytically provided that $\gamma$ is fixed \cite{achiam2017constrained}; $
	{D}_{KL}\left({{\mathit{\pi}}_{\boldsymbol{\theta}{\mathbf{'}}}\left({{\mathbf{s}}_{n}}\right)\parallel{\mathit{\pi}}_{\boldsymbol{\theta}}\left({{\mathbf{s}}_{n}}\right)}\right)
	$ represents the Kullback-Leibler (KL) divergence between the new and old policy and is given by 
	\[
	{D}_{KL}\left({{\mathit{\pi}}_{\boldsymbol{\theta}{\mathbf{'}}}\parallel{\mathit{\pi}}_{\boldsymbol{\theta}}}\right){\triangleq}\mathop{\mathbb{E}}\limits_{{\mathbf{a}}_{n}\sim{\mathit{\pi}}_{\boldsymbol{\theta}{\mathbf{'}}}}\left[{\log\frac{{\mathit{\pi}}_{\boldsymbol{\theta}{'}}({\mathbf{a}}_{n}|{\mathbf{s}}_{n})}{{\mathit{\pi}}_{\boldsymbol{\theta}}({\mathbf{a}}_{n}|{\mathbf{s}}_{n})}}\right]
	\]
	The equality in (\ref{bound1}) follows from the definition of the advantage function in (\ref{advantage}) and algebraic manipulation; the equality in (\ref{bound2}) follows since we utilize importance sampling to change the expectation to be over actions sampled from ${\mathit{\pi}}_{\boldsymbol{\theta}}$ instead of ${\mathit{\pi}}_{\boldsymbol{\theta}{'}}$. The inequality in (\ref{bound3}) follows from utilizing a finite-time horizon in Corollary 2 in \cite{achiam2017constrained}. Furthermore, we observe that by setting $\boldsymbol{\theta'}=\boldsymbol{\theta}$, both the difference in expected reward and the lower bound equal zero. Thus, we conclude that the given lower bound minorizes the difference in discounted rewards, and optimizing this lower bound while ensuring that it is nonnegative, we can always find a value of the policy parameters that does not lead to a decrease in expected reward.
	
	The lower bound in (\ref{bound3}) is useful because it allows us to bound the improvement in the expected reward in terms of the change to the policy parameters via the KL-divergence; this is in sharp contrast to the REINFORCE algorithm in which the effect of taking too large a step cannot be quantified. At the same time, directly optimizing this lower bound leads to conservatively small step sizes and impractically long training times \cite{schulman2015trust}. Instead, following \cite{achiam2017constrained}, we consider the following proxy optimization problem:}
	\begin{subequations}
		\begin{align}\label{surrogate_problem}
		{\mathop{\mathrm{maximize}}\limits_{\boldsymbol{\theta'}}\hspace{1.33em}{\mathcal{L}}_{\boldsymbol{\theta}}\left({\boldsymbol{\theta}{\mathbf{'}}}\right)}\hspace{7.60em}\\
		\mathrm{subject}\hspace{0.33em}\mathrm{to}\hspace{0.33em}{\mathop{\mathbb{E}}\limits_{{{\mathbf{s}_{n'}\sim{d}_{{\mathit{\pi}}_{\boldsymbol{\theta}}}}\atop{}}}\left[{{D}_{KL}\left({{\mathit{\pi}}_{\boldsymbol{\theta}{'}}\parallel{\mathit{\pi}}_{\boldsymbol{\theta}}}\right)}\right]\leq\mathit{\delta}}\label{KL_constraint}
		\end{align}
	\end{subequations}
    In its current form, this problem is non-convex and thus mathematically intractable. However, the objective function and constraint in (\ref{surrogate_problem}) can be approximated using their Taylor series expansion as follows:
	\begin{subequations}
		\begin{align}
		{\mathcal{L}}_{\boldsymbol{\theta}}\left({\boldsymbol{\theta}{\mathbf{'}}}\right)\approx{\mathcal{L}}_{\boldsymbol{\theta}}\left({\boldsymbol{\theta}}\right){+}{\mathbf{g}}_{\boldsymbol{\theta}}^{{T}}\mathbf{\left({\boldsymbol{\theta}{'}{-}\boldsymbol{\theta}}\right)}\hspace{6.7em}\\
		\mathop{\mathbb{E}}\limits_{{{\mathbf{s}_{n'}\sim{d}_{{\mathit{\pi}}_{\boldsymbol{\theta}}}}\atop{}}}\left[{{D}_{KL}\left({{\mathit{\pi}}_{\boldsymbol{\theta}{'}}\parallel{\mathit{\pi}}_{\boldsymbol{\theta}}}\right)}\right]\approx\frac{1}{2}{\left({\boldsymbol{\theta}{\mathbf{'}}\mathbf{{-}}\boldsymbol{\theta}}\right)}^{T}{\mathbf{F}}_{\boldsymbol{\theta}}\left({\boldsymbol{\theta}{\mathbf{'}}\mathbf{{-}}\boldsymbol{\theta}}\right)
		\end{align}
	\end{subequations}
	where
	\begin{equation}
	\begin{gathered}
		{\mathbf{F}_{\boldsymbol{\theta}}}{\triangleq}{\nabla}_{\boldsymbol{\theta}{'}}^{2}\mathop{\mathbb{E}}\limits_{{{{\mathbf{s}}_{t'}\sim{d}_{{\mathit{\pi}}_{\boldsymbol{\theta}}}}\atop{}}}\left[{{D}_{KL}\left({{\mathit{\pi}}_{\boldsymbol{\theta}{'}}\parallel{\mathit{\pi}}_{\boldsymbol{\theta}}}\right)}\right]{|}_{\boldsymbol{\theta}}\hspace{0em}{=}\mathop{\mathbb{E}}\limits_{{{{\mathbf{a}}_{n'}\sim{\mathit{\pi}}_{\boldsymbol{\theta}}}\atop{{\mathbf{s}}_{n'}\sim{d}_{{\mathit{\pi}}_{\boldsymbol{\theta}}}}}}\left[{{\nabla}_{\boldsymbol{\theta}{\mathbf{'}}}\log{\mathit{\pi}}_{\boldsymbol{\theta}{'}}\left({{\mathbf{a}}_{n'}|{\mathbf{s}}_{n'}}\right){|}_{\boldsymbol{\theta}}{\nabla}_{\boldsymbol{\theta}{\mathbf{'}}}\log{\mathit{\pi}}_{\boldsymbol{\theta}{'}}\left({{\mathbf{a}}_{n'}|{\mathbf{s}}_{n'}}\right){|}_{\boldsymbol{\theta}}^{T}}\right]\hspace{0.80em}\label{F_theta}
	\end{gathered}
\end{equation}
	We remark at this juncture that $\mathbf{F}_{\boldsymbol{\theta}}$ can be interpreted as the Fisher Information matrix (FIM) of the policy and describes the curvature as a function of the policy parameter. Utilizing these approximations to both the objective function and constraints, we obtain the following convex proxy optimization problem:
	\begin{subequations}
		\begin{align}
		\mathop{\mathrm{maximize}}\limits_{\mathit{\boldsymbol{\theta'}}}\hspace{0.33em}{\mathbf{g}}_{\boldsymbol{\theta}}^{{T}}\left({\boldsymbol{\theta}{\mathbf{'}}\mathbf{{-}}\boldsymbol{\theta}}\right)\hspace{6.05em}\\
		\mathrm{subject}\hspace{0.33em}\mathrm{to}\hspace{0.33em}\frac{1}{2}{\left({\boldsymbol{\theta}{\mathbf{'}}\mathbf{{-}}\boldsymbol{\theta}}\right)}^{T}{\mathbf{F}_{\boldsymbol{\theta}}}\left({\boldsymbol{\theta}{\mathbf{'}}\mathbf{{-}}\boldsymbol{\theta}}\right)\leq\mathit{\delta}\label{second_order_KL_constraint}
		\end{align}
	\end{subequations}
	This proxy problem in (\ref{surrogate_problem}) can be solved analytically to yield the optimal value of $\boldsymbol{\theta'}$ in terms of the current parameter $\boldsymbol{\theta}$ as:
	\begin{equation}\label{update_direction}
	\mathit{\boldsymbol{\theta}}{'}{=}\mathit{\boldsymbol{\theta}}{+}\sqrt{\frac{{2}\delta}{{\mathbf{g}}_{\boldsymbol{\theta}}^{T}{\mathbf{F}}_{\boldsymbol{\theta}}^{{-}{1}}{\mathbf{g}}_{\boldsymbol{\theta}}}}{\mathbf{F}}_{\boldsymbol{\theta}}^{{-}{1}}{\mathbf{g}}_{\boldsymbol{\theta}}
	\end{equation}
	We observe that the proxy optimization problem requires knowledge of ${\mathbf{g}}_{\boldsymbol{\theta}}$ and ${\mathbf{F}}_{\boldsymbol{\theta}}$; however, we are unable to evaluate them analytically. This can be overcome as follows: we sample $M$ episodes ${\mathit{\varepsilon}}^{1}{,}\ldots{,}{\mathit{\varepsilon}}^{M}$ from the current policy ${\mathit{\pi}}_{{\boldsymbol{\theta}}_{}}$, where:
	\[{\mathit{\varepsilon}}^{m}{=}\left\{{{\mathbf{s}}_{1}^{{m}}{,}{\mathbf{a}}_{1}^{{m}}{,}{r}_{1}^{m}{,}\ldots{,}{\mathbf{s}}_{N}^{{m}}{,}{\mathbf{a}}_{N}^{{m}}{,}{r}_{N}^{m}}\right\}\]
	
	These episodes can be used to obtain unbiased estimates of the policy gradient and FIM as follows:
	\begin{equation}\label{g_estimate}
	\hat{\mathbf{g}}_{\mathbf{\boldsymbol{\theta}}}{=}\frac{1}{MN}\mathop{\sum}\limits _{{m}{=}{1}}^{M}{\mathop{\sum}\limits _{{n}{'}{=}{1}}^{N}{{\mathit{\gamma}}^{n'}{\nabla}_{\boldsymbol{\theta'}}\log{\mathit{\pi}}_{\boldsymbol{\theta'}}\left({{\mathbf{a}}_{n'}^{{m}}|{\mathbf{s}}_{n'}^{{m}}}\right)}}{\hat{A}}^{{\mathit{\pi}}_{\boldsymbol{\theta'}}}{(}{\mathbf{s}}_{n'}^{{m}}{,}{\mathbf{a}}_{n'}^{{m}}{)}{|}_{\boldsymbol{\theta'}\mathbf{{=}}{\boldsymbol{\theta}}_{{}}}
	\end{equation}
	
	\begin{equation}\label{F_estimate}
	\hat{\mathbf{F}}_{\boldsymbol{\theta}}{=}\frac{1}{MN}\mathop{\sum}\limits_{{m}{=}{1}}\limits^{M}{\mathop{\sum}\limits_{{n'}{=}{1}}\limits^{N}{{[}{\nabla}_{\boldsymbol{\theta}{\mathbf{'}}}\log{\mathit{\pi}}_{\boldsymbol{\theta}{'}}\left({{\mathbf{a}}_{n'}^{{m}}|{\mathbf{s}}_{n'}^{{m}}}\right)\cdot}}
	{\nabla}_{\boldsymbol{\theta}{\mathbf{'}}}\log{\mathit{\pi}}_{\boldsymbol{\theta}{'}}{\left({{\mathbf{a}}_{n'}^{{m}}|{\mathbf{s}}_{n'}^{{m}}}\right)}^{T}]{|}_{\boldsymbol{\theta}{\mathbf{'}}\mathbf{{=}}\boldsymbol{\theta}}
	\end{equation}
%	\newpage
    {\color{black}
    While directly utilizing the update in (\ref{update_direction}) is possible \cite{kakade2002natural}, it can lead to a violation of the KL-divergence constraint in (\ref{KL_constraint}) since we utilize Taylor approximations rather than the original objective function and constraints. To overcome this issue, we use a backtracking line search with decaying step sizes as proposed in \cite{achiam2017constrained}; thus, intead in (\ref{update_direction}) denoting:
	\begin{equation}\label{delta}
	{\Delta}_{\boldsymbol{\theta}}{=}\sqrt{\frac{{2}\delta}{\hat{\mathbf{g}}_{\boldsymbol{\theta}}^{T}\hat{\mathbf{F}}_{\boldsymbol{\theta}}^{{-}{1}}\hat{\mathbf{g}}_{\boldsymbol{\theta}}}}\hat{\mathbf{F}}_{\boldsymbol{\theta}}^{{-}{1}}\hat{\mathbf{g}}_{\boldsymbol{\theta}}
	\end{equation}
	we use the following parameter update
	\begin{equation}\label{correct_update}		\boldsymbol{\theta}{'}{=}\boldsymbol{\theta}{+}{\zeta}^{\mathit{j}}{\Delta}_{\mathit{\boldsymbol{\theta}}}
	\end{equation}
	instead of the one in (\ref{update_direction}), where $\zeta\in(0,1)$ is the step size and the integer $j=0,1,\ldots$ is incremented (thus decaying the step size by a factor of $\zeta$ with each increment) until the KL-divergence constraint is satisfied and the objective function from (\ref{L}) is nonnegative, i.e.,}
	\begin{equation}\label{L_nonnegative}
	{\mathcal{L}}_{\boldsymbol{\theta}}\left({\boldsymbol{\theta}{\mathbf{'}}}\right)\geq{0}
	\end{equation}
	\begin{equation}\label{KL_constraint_respected}
	\mathop{\mathbb{E}}\limits_{{{\mathbf{s}_{t'}\sim{d}_{{{\pi}}_{\boldsymbol{\theta}}}}\atop{}}}\left[{{D}_{KL}\left({{\mathit{\pi}}_{\boldsymbol{\theta}{'}}\parallel{\mathit{\pi}}_{\boldsymbol{\theta}}}\right)}\right]\leq\mathit{\delta}
	\end{equation}
	
	Finally, combining all these steps together, we obtain the deep trust-region reinforcement learning algorithm in Algorithm \ref{alg1}. We emphasize at this point that the value network is necessary only during the training phase; once the centralized policy has been trained using Algorithm \ref{alg1}, the current state can be directly forward-propagated through the policy network to obtain the network power allocation. 
	
	\begin{algorithm}[h!]
		\caption{Trust Region Policy Optimization for Centralized Power Allocation}\label{alg1}
		\begin{algorithmic}[1]
			\State {\textbf{initialize} centralized policy and value network parameters $\boldsymbol{\theta}_{0}$, $\boldsymbol{\phi}_{0}$.}
			\State {\textbf{for} $i=0,1,\ldots,N_{iterations}$ \textbf{do}}			
			\State {\hspace{1.4em}\textbf{for} $m=0,1,\ldots,M$ \textbf{do}}
			\State {\hspace{3em} Sample  ${\mathit{\pi}}_{{\boldsymbol{\theta}}_{i}}$ to collect ${\varepsilon}^{m}{=}{\left\{{{\mathbf{s}}_{1}^{\mathit{m}}{,}{\mathbf{a}}_{1}^{\mathit{m}}{,}{R}^{m}\left({{\mathbf{H}}^{(1)},{\mathbf{P}}^{(1)}}\right){,}\ldots{,}{\mathbf{s}}_{N}^{\mathit{m}}{,}{\mathbf{a}}_{N}^{\mathit{m}}{,}{R}^{m}\left({{\mathbf{H}}^{(n)},{\mathbf{P}}^{(n)}}\right)}\right\}}$.}
			\State \hspace{1em} \textbf{end for}		
			\State {\hspace{1em} Estimate ${{A}}_{}^{{\mathit{\pi}}_{{\boldsymbol{\theta}}_{i}}}$.}
			\State {\hspace{1em} Update critic network parameter $\boldsymbol{\phi}_{i}$ to fit estimated advantage values.}
			\State {\hspace{1em} Compute policy gradient estimate ${\mathbf{g}}_{\boldsymbol{\theta}_{i}}$ using (\ref{g_estimate}).}
			\State {\hspace{1em} Compute FIM estimate $\hat{\mathbf{F}}_{\boldsymbol{\theta}_{i}}$ using (\ref{F_estimate})}.
			\State {\hspace{1em} Compute policy update direction $\Delta_{i}$ using (\ref{delta}).}
			\State {\hspace{1em} \textbf{for} $j=0,1,\ldots$ \textbf{do}}
			\State \hspace{2em} Compute: \[{\boldsymbol{\theta}_{i+1}}_{{{}}{{}}{{}}}{=}{\boldsymbol{\theta}}_{{i}}{+}{\mathit{\zeta}}^{j}{\Delta}_{i}\]
			\State \hspace{2em} \textbf{if} (\ref{L_nonnegative}) and (\ref{KL_constraint_respected}) are satisfied \textbf{then}
			\State \hspace{3em} break
			\State \hspace{2em} \textbf{end if}		
			\State \hspace{1em} \textbf{end for}		
			\State \hspace{0em}\textbf{end for}
		\end{algorithmic}
	\end{algorithm}
	\subsection{Decentralized Multi-Agent Approaches}
	{\color{black}
	The centralized single-agent approach we have developed so far has the same information exchange requirements as the FP and WMMSE algorithms, which become burdensome as the number of cells and users increases. More importantly, the increasing size of both the state and action spaces leads to extremely slow convergence. This issue, commonly referred to as the `curse-of-dimensionality' \cite{zahavy2018learn,han2018solving}, renders the centralized DRL approach unsuitable for large wireless networks as the associated training times become impractically long.
	
	Multi-agent approaches are attractive from a computational perspective since the policy network only computes the actions for a single agent; this circumvents the problem of the increasing size of the state and action spaces that is inevitably incurred with a centralized approach. Furthermore, as shown in Figures \ref{single_agent_DRL} and \ref{multi_agent_DRL}, each agent processes only its own state information, but can learn under a \textit{common reward}. Intuitively, training a single policy for deployment across all BSs allows the network size to grow without necessarily increasing the training time required. As a result, multi-agent approaches are generally utilized to solve large-scale reinforcement learning problems as opposed to single-agent methods \cite{baker2019emergent,omidshafiei2017deep,foerster2016learning}. However, the trust region methods employed in \cite{achiam2017constrained} are developed for a centralized single-agent approach.
%{\color{blue}
%    \begin{figure}[t!]
%	\centering
%	\begin{subfigure}[]{0.45\textwidth}
%		\centering
%		\includegraphics[trim={4cm 6cm 9cm 0cm},clip, height=0.70\textwidth]{single_agent_DRL.pdf}
%		\caption{Single-agent DRL.}\label{single_agent_DRL}
%	\end{subfigure}\hfill
%	\begin{subfigure}[]{0.45\textwidth}
%		\centering
%		\includegraphics[trim={4cm 6cm 9cm 0cm},clip, height=0.70\textwidth]{multi_agent_DRL.pdf}
%		\caption{Multi-agent DRL}\label{multi_agent_DRL}
%	\end{subfigure}
%
%	\caption{{\color{blue}In multi-agent DRL methods, each agent acts individually. The collective actions of all agents influence the common reward and the (usually distinct) states the agents will encounter in the next time step.}}
%\end{figure} 
%}
{\color{black}
	\begin{figure}[t!]
		\begin{subfigure}[]{0.45\textwidth}
			\includegraphics[trim={0cm 0cm 0cm 01cm},clip, height=0.60\textwidth]{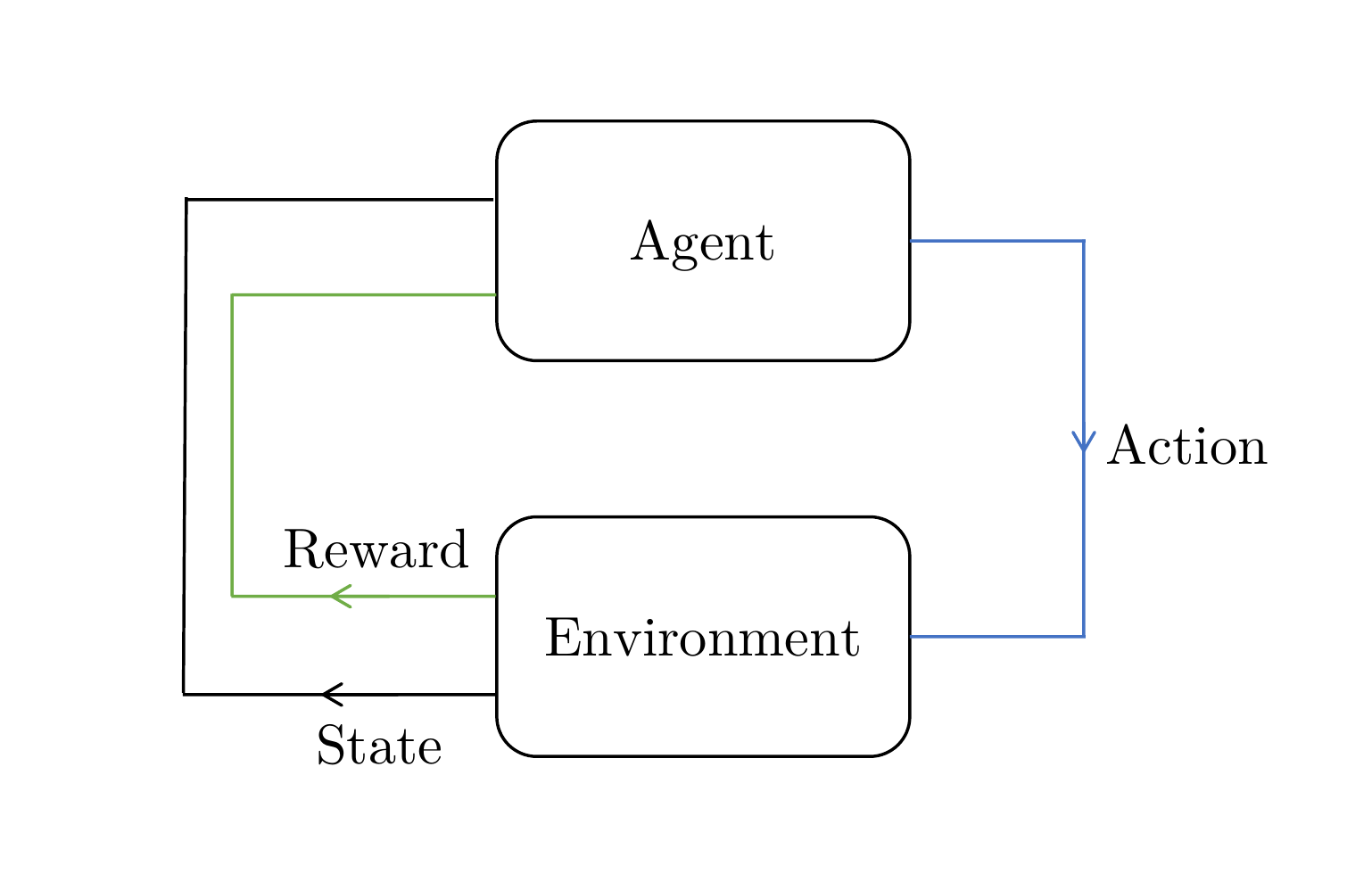}
			\caption{Single-agent DRL.}\label{single_agent_DRL}
		\end{subfigure}\hfill
		\begin{subfigure}[]{0.45\textwidth}
			\hspace{-4.00em}
			\includegraphics[trim={0cm 0cm 0cm 0cm},clip, height=0.58\textwidth]{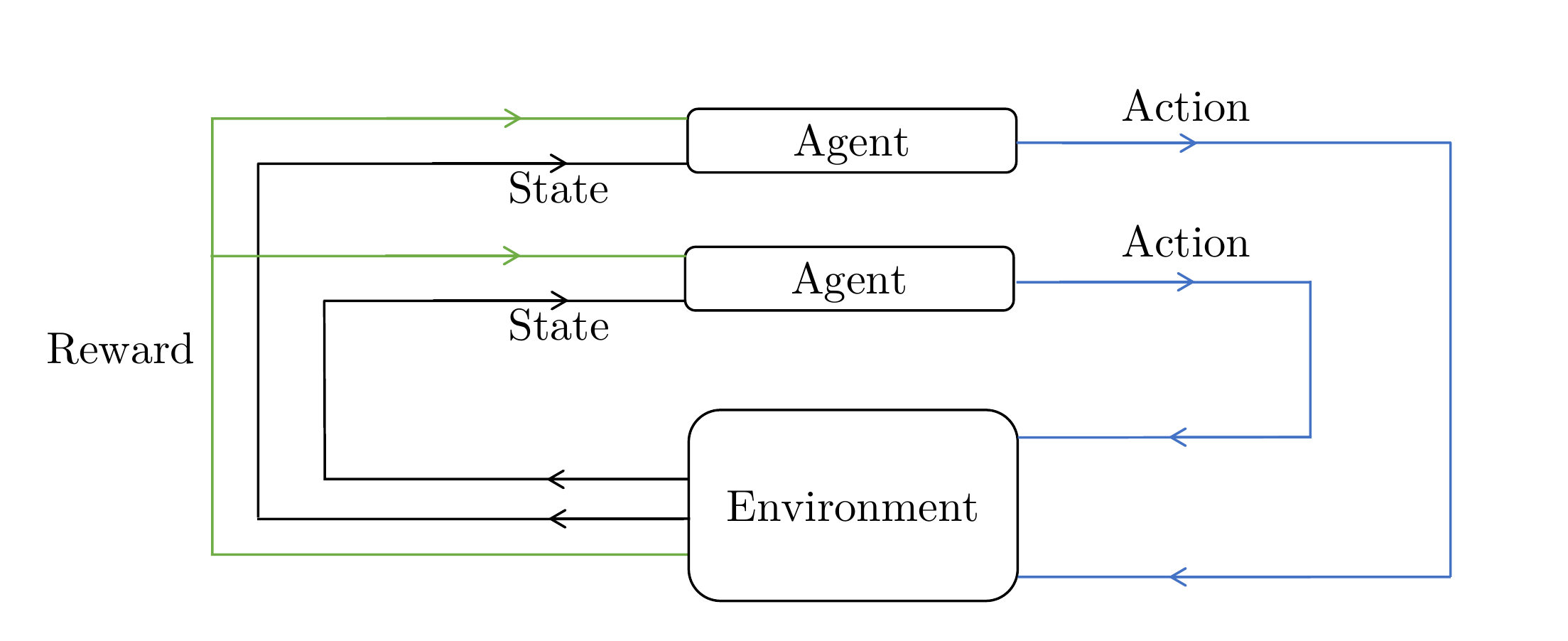}
			\caption{Multi-agent DRL}\label{multi_agent_DRL}
		\end{subfigure}
		
		\caption{{\color{black}In multi-agent DRL methods, each agent acts individually. The collective actions of all agents influence the common reward and the (usually distinct) states the agents will encounter in the next time step.}}
	\end{figure} 
}

	We overcome these challenges by modifying the trust region policy optimization approach to function in a decentralized fashion with limited information exchange. More specifically, we consider each BS in the network to be an individual agent with partial access to network CSI; however, we design the learning process so that these agents learn under the \textit{common reward} of \textit{network spectral efficiency}. This is in contrast to prior DRL approaches explored in the literature, since methods like REINFORCE are typically not robust enough to be used in a multi-agent setting, due to the aforementioned sensitivity to changes in the policy parameters. While deep Q-learning has been used in multi-agent settings, as mentioned earlier, it cannot directly optimize the network spectral efficiency and requires significant feature engineering. On the other hand, constrained policy optimization algorithms like trust region policy optimization have been successfully employed in solving multi-agent common-objective learning problems \cite{schulman2015trust,baker2019emergent}.
	 
	\textit{Partially Decentralized Approach}: Accordingly, we propose a partially decentralized multi-agent approach with reduced information exchange and per-BS CSI requirements. We propose a round-robin approach in which the BSs sequentially allocate powers, in a randomly chosen order, to the users within their cells, using an identical policy network deployed at each BS. Thus, the time horizon for the multi-agent approach is $N=B$. To limit the information exchange, each BS is allowed access to its own downlink CSI (but not the downlink CSI of other BSs) as well as the power allocation information of the previous BSs. Thus, the state and action for BS $b$ are given by:
	\[
	{\mathbf{s}}_{b}{=}\left\{{\left\{{{h}_{{b}\rightarrow{k}{,}{b}{'}}^{\left({n}\right)}{|}{b}{'}{=}{1}{,}\ldots{,}{B}{;}\hspace{0.33em}{k}{=}{1}{,}\ldots{,}{K}}\right\}{,}\left\{{{p}_{{b}{'}\rightarrow{k}}^{\left({n}\right)}{|}{b}{'}{=}{1}{,}\ldots{,}{b}{-}{1}{;}\hspace{0.33em}{k}{=}{1}{,}\ldots{,}{K}}\right\}}\right\}
	\]
	\[
	{\mathbf{a}}_{b}{=}\left\{{{p}_{{b}\rightarrow{k}}^{\left({n}\right)}{|}{k}{=}{1}{,}\ldots{,}{K}}\right\}\hspace{23em}
	\]
	
	To clarify, in this round-robin allocation scheme, the initial power allocation is set to zero, and BS $1$ allocates power to the users within its own cell using its policy output based on its downlink CSI. It then forwards its power allocation to BS $2$, which then uses this power information along with its own downlink CSI to allocate power to its own users \textit{using the same policy as deployed at BS $1$}. This process is halted once every BS in the network has allocated power to its users, as illustrated in Figure $\ref{decentralized_diagram}$. 

	To avoid notational clutter in our subsequent derivations, we do not explicitly indicate the randomized order of BSs, and assume a single-time slot episode length for the decentralized methods\footnote{Note that an episode can alternatively be defined to comprise multiple time slots; however, in this work we consider a single time-slot per episode without loss of generality.}. We remark that this procedure leads to slight differences in the definition of an episode: for the centralized setting, an episode can encompass multiple time slots, but for the partially decentralized case we consider an episode to be complete once all BSs have allocated the transmit powers to users for that particular time slot. Additionally, the power allocation values are sampled from a normal distribution dependent on the policy network output similar to the centralized setting. 
	
	{\color{black}
    Unlike the centralized approach, however, the probability of transition to a new state does depend upon the previous action since an \textit{identical policy} is utilized at each BS. Specifically, each BS's state consists of its own downlink CSI and the power allocated by the previous BSs; the former is independent of the previous action as in the centralized approach, while the latter directly depends upon the previous action chosen by the policy. With this crucial difference, the actions of all BSs are coupled; and the problem can now be considered in the more general MDP framework as introduced in Section \textrm{III} earlier with multiple time-steps. We remark once again that this decentralized setting is similar to the game-theoretic approaches detailed earlier \cite{menache2011network,park2010distributed} as well as prior works that have utilized DRL for solving resource allocation problems \cite{de2018team}.}}
	
	 Allowing each BS access to its downlink CSI is obviously necessary; the additional power allocation information of the previous BSs should also allow it to avoid creating intercell interference. For instance, if the power allocated by BS $b$ to user $k$ is high, BS $b+1$ should avoid creating interference to that user on channel ${h}_{{b}{+}{1}\rightarrow{k}{,}{b}}$ with its own power allocation strategy since that could potentially reduce the network sum-rate. 
	 
	 {\color{black} We emphasize that such an approach with partial information sharing is only possible in the framework of multi-agent reinforcement learning; model-based optimization algorithms like WMMSE and FP require full exchange of CSI between all BSs to enable cooperative power allocation. Also, the prior decentralized approaches in the literature do not allow for flexible information sharing: for example, the sequential sharing of power information by BSs in our proposed partially decentralized algorithm would simply not be possible in either the game-theoretic decentralized approaches presented in \cite{menache2011network,de2015best} or the optimization approaches presented in \cite{tervo2017distributed,fritzsche2013distributed,park2010distributed}, as this information would be useless without the corresponding downlink channel state information of other BSs.}

    \begin{figure}[t!]
	\centering
	\begin{subfigure}[]{0.5\textwidth}
		\centering
		\includegraphics[trim={5cm 0cm 0cm 0cm},clip, height=0.9\textwidth]{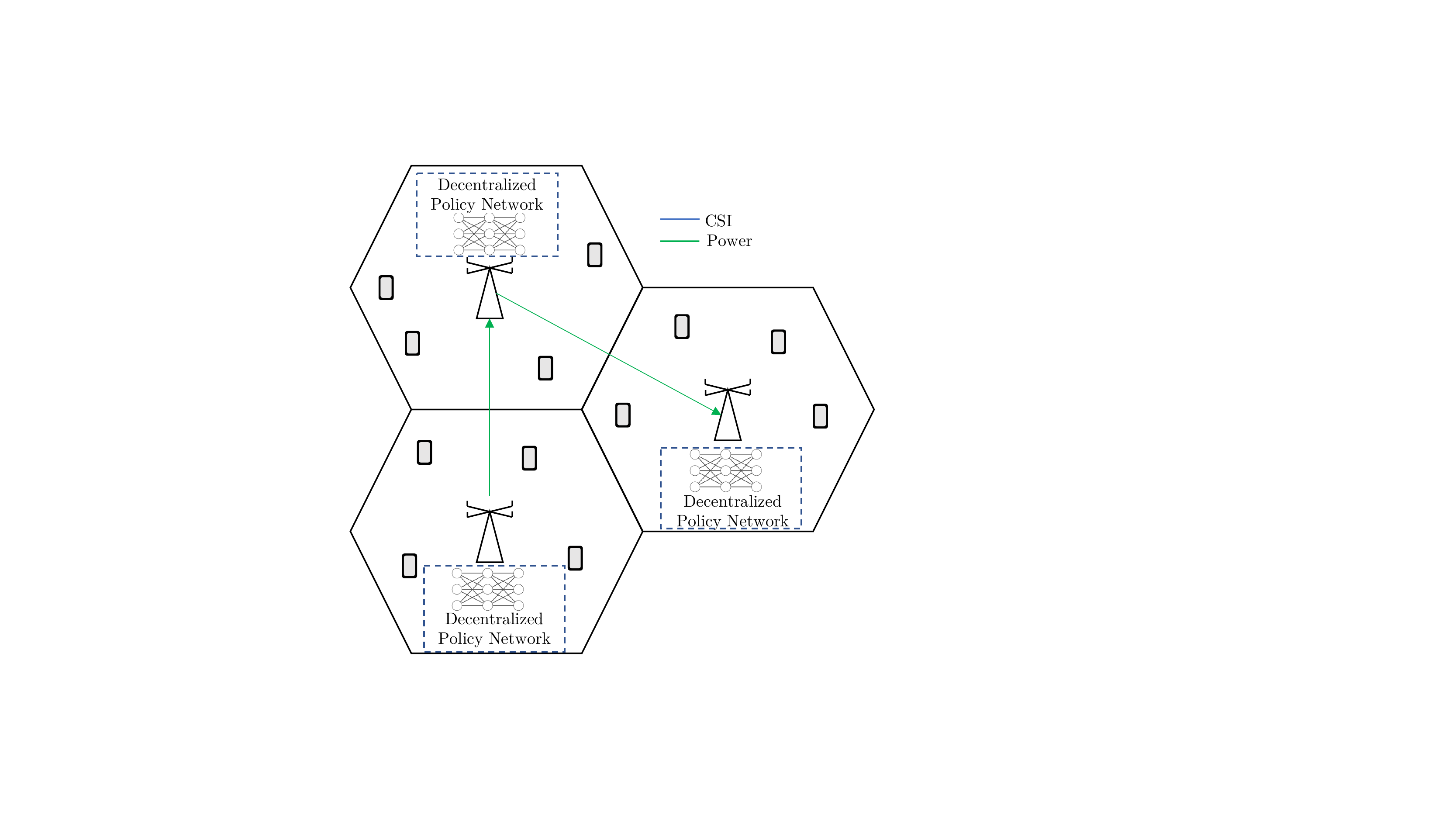}
		\caption{}\label{decentralized_diagram}
	\end{subfigure}\hfill
    \begin{subfigure}[]{0.5\textwidth}
		\centering
		\includegraphics[trim={5cm 0cm 0cm 0cm},clip, height=0.9\textwidth]{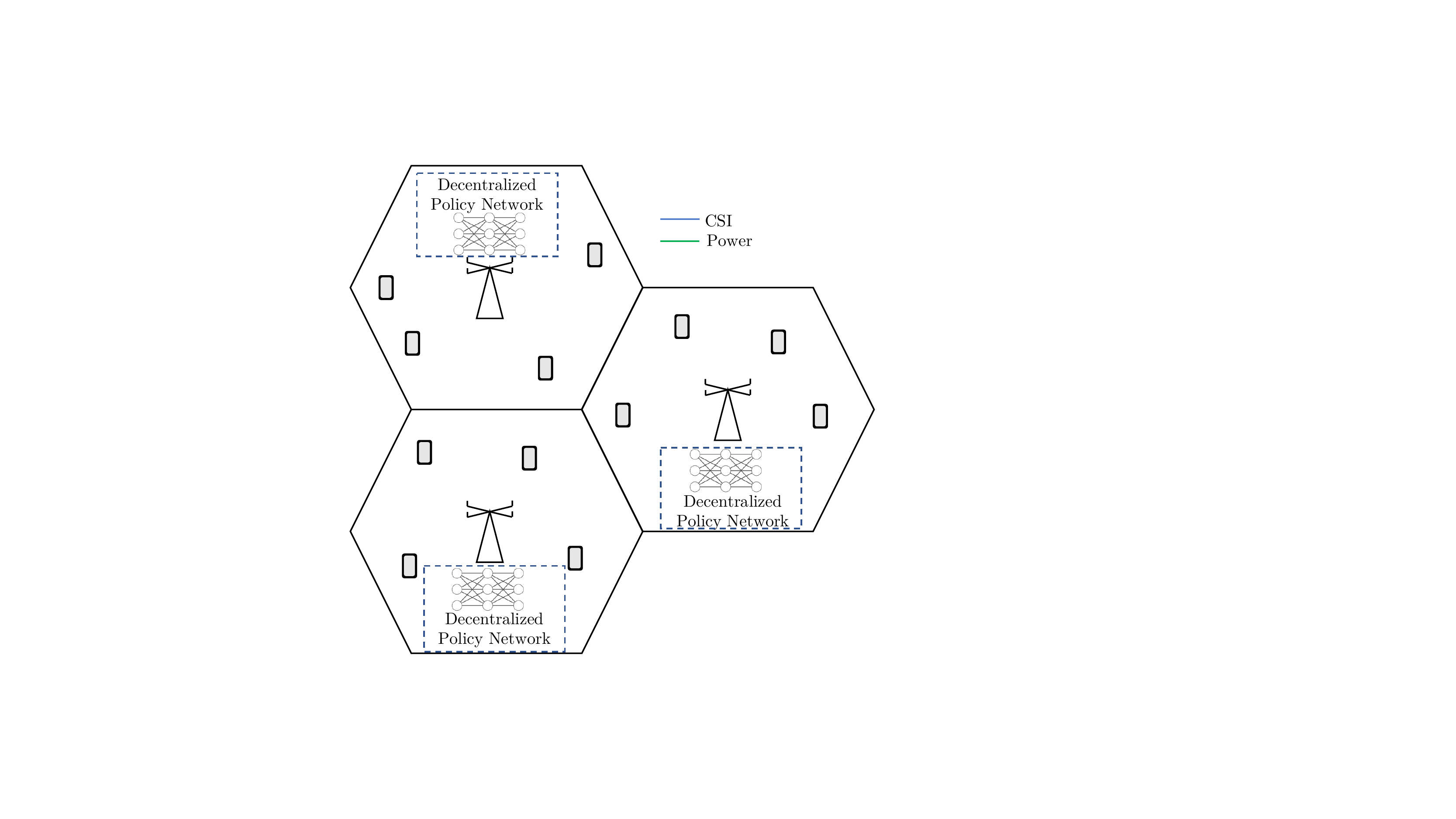}
		\caption{}\label{decentralized_diagrams}
	\end{subfigure}
	\caption{For the partially decentralized approach, the BSs sequentially exchange power allocation information in a predetermined order until the last BS is reached. For the fully decentralized approach, no CSI or power information is exchanged, and the BSs simultaneously and independently determine their power allocation strategy.}
\end{figure} 

    As with the centralized approach, the reward function is chosen to be the network sum spectral efficiency; however, to ensure that all BSs learn under a common reward, the reward is accessible only at the end of each episode (i.e., once all BSs have allocated powers to their associated users) and intermediate values of reward are set to zero. Hence, a complete episode is given by:
    \[
    \varepsilon{=}{\left\{{{\mathbf{s}}_{1}{,}{\mathbf{a}}_{1}{,}{0}{,}\ldots{,}{\mathbf{s}}_{\mathrm{B}}{,}{\mathbf{a}}_{B}{,}{R}\left({{\mathbf{H}}^{(n)},{\mathbf{P}}^{(n)}}\right)}\right\}}
    \]
    The ascent direction is once again calculated using (\ref{delta}). Since an identical policy is deployed at all BSs, the same gradient update is applied to every agent. Combining all steps together, the partially decentralized approach is summarized in Algorithm \ref{alg2}.
	\begin{algorithm}[h!]
	\caption{Trust Region Policy Optimization for Partially Decentralized Power Allocation}\label{alg2}
	\begin{algorithmic}[1]
		\State {\textbf{initialize} policy and value network parameters $\boldsymbol{\theta}_{0}$, $\boldsymbol{\phi}_{0}$ at BSs $b=1,\ldots,B$.}
		\State {\textbf{for} $i=0,1,\ldots,N_{iterations}$ \textbf{do}}			
		\State {\hspace{1.4em}\textbf{for} $m=0,1,\ldots,M$ \textbf{do}}		
		\State {\hspace{2.3em}\textbf{for} $b=0,1,\ldots,B$ \textbf{do}}
		\State {\hspace{4.2em} Propagate ${\mathbf{s}}_{b}^{\mathit{m}}$ through ${\mathit{\pi}}_{{\boldsymbol{\theta}}_{i}}$ at BS $b$ to obtain ${\mathbf{a}}_{b}^{\mathit{m}}$.}
		\State {\hspace{4.2em} Forward ${\mathbf{a}}_{1}^{\mathit{m}},\ldots,{\mathbf{a}}_{b}^{\mathit{m}}$ to BS $b+1$.}
		\State \hspace{2em} \textbf{end for}
		\State {\hspace{2em} Collect ${\varepsilon}^{m}{=}{\left\{{{\mathbf{s}}_{1}^{\mathit{m}}{,}{\mathbf{a}}_{1}^{\mathit{m}}{,}{0}{,}\ldots{,}{\mathbf{s}}_{B}^{\mathit{m}}{,}{\mathbf{a}}_{B}^{\mathit{m}}{,}{R}^{m}\left({{\mathbf{H}}^{(n)},{\mathbf{P}}^{(n)}}\right)}\right\}}$.}				
		\State \hspace{1em} \textbf{end for}		
		\State {\hspace{1em} Estimate ${{A}}_{}^{{\mathit{\pi}}_{{\boldsymbol{\theta}}_{i}}}$.}
		\State {\hspace{1em} Update critic network parameter $\boldsymbol{\phi}_{i}$ at BSs $b=1,\ldots,B$ to fit estimated advantage values.}
		\State {\hspace{1em} Compute policy gradient estimate ${\mathbf{g}}_{\boldsymbol{\theta}_{i}}$ using (\ref{g_estimate}).}
		\State {\hspace{1em} Compute FIM estimate $\hat{\mathbf{F}}_{\boldsymbol{\theta}_{i}}$ using (\ref{F_estimate})}.
		\State {\hspace{1em} Compute policy update direction $\Delta_{i}$ using (\ref{delta}).}
		\State {\hspace{1em} \textbf{for} $j=0,1,\ldots$ \textbf{do}}
		\State \hspace{2em} Compute: \[{\boldsymbol{\theta}_{i+1}}_{{{}}{{}}{{}}}{=}{\boldsymbol{\theta}}_{{i}}{+}{\mathit{\zeta}}^{j}{\Delta}_{i}\]
		\State \hspace{2em} \textbf{if} (\ref{L_nonnegative}) and (\ref{KL_constraint_respected}) are satisfied \textbf{then}
		\State \hspace{3em} Apply policy update at BSs $b=1,\ldots,B$.
		\State \hspace{3em} break
		\State \hspace{2em} \textbf{end if}		
		\State \hspace{1em} \textbf{end for}		
		\State \hspace{0em}\textbf{end for}
	\end{algorithmic}
	\end{algorithm}

    Similar to the centralized setting, we remark that once the (identical) policy network at each BS has been trained, the value network is no longer necessary. Likewise, the common gradient update needs to be applied only during the training phase; once the trained agents are deployed, the state information can directly be propagated through the policy network at each BS to obtain the desired power allocation for that cell.
    
    \textit{Fully Decentralized Approach:} For the fully decentralized approach, we eliminate all communication between the BSs in the network; hence the state information for each BS comprises only its downlink CSI:
    \[
    {\mathbf{s}}_{b}{=}\left\{{{h}_{{b}\rightarrow{k}{,}{b}{'}}^{\left({n}\right)}{|}{b}{'}{=}{1}{,}\ldots{,}{B}{;}\hspace{0.33em}{k}{=}{1}{,}\ldots{,}{K}}\right\}
    \]
    \[
    {\mathbf{a}}_{b}{=}\left\{{{p}_{{b}\rightarrow{k}}^{\left({n}\right)}{|}{k}{=}{1}{,}\ldots{,}{K}}\right\}\
    \]
    
    Since no information is exchanged between the agents, all BSs independently and simultaneously determine their individual power allocation strategies, as illustrated in Figure $\ref{decentralized_diagrams}$. The reward function is once again chosen to be the \textit{network sum-rate} and is only accessible at the end of each episode; likewise, an identical policy is utilized at each BS and updated at the end of each training episode. The fully decentralized algorithm is summarized in Algorithm $\ref{alg3}$.
    
    Finally, we note that the feedback of the network sum-rate to both the centralized agent and the distributed agents is only necessary during the training phase. Once a trained agent has been deployed, there is no further need for this information.
    
    \begin{algorithm}[h!]
    	\caption{Trust Region Policy Optimization for Fully Decentralized Power Allocation}\label{alg3}
    	\begin{algorithmic}[1]
    		\State {\textbf{initialize} policy and value network parameters $\boldsymbol{\theta}_{0}$, $\boldsymbol{\phi}_{0}$ at BSs $b=1,\ldots,B$.}
    		\State {\textbf{for} $i=0,1,\ldots,N_{iterations}$ \textbf{do}}			
    		\State {\hspace{1.4em}\textbf{for} $m=0,1,\ldots,M$ \textbf{do}}		
    		\State {\hspace{2.3em}\textbf{for} $b=0,1,\ldots,B$ \textbf{do}}
    		\State {\hspace{4.2em} Propagate ${\mathbf{s}}_{b}^{\mathit{m}}$ through ${\mathit{\pi}}_{{\boldsymbol{\theta}}_{i}}$ at BS $b$ to obtain ${\mathbf{a}}_{b}^{\mathit{m}}$.}
    		\State \hspace{2em} \textbf{end for}
    		\State {\hspace{2em} Collect ${\varepsilon}^{m}{=}{\left\{{{\mathbf{s}}_{1}^{\mathit{m}}{,}{\mathbf{a}}_{1}^{\mathit{m}}{,}{0}{,}\ldots{,}{\mathbf{s}}_{B}^{\mathit{m}}{,}{\mathbf{a}}_{B}^{\mathit{m}}{,}{R}^{m}\left({{\mathbf{H}}^{(n)},{\mathbf{P}}^{(n)}}\right)}\right\}}$.}				
    		\State \hspace{1em} \textbf{end for}		
    		\State {\hspace{1em} Estimate ${{A}}_{}^{{\mathit{\pi}}_{{\boldsymbol{\theta}}_{i}}}$.}
    		\State {\hspace{1em} Update critic network parameter $\boldsymbol{\phi}_{i}$ at BSs $b=1,\ldots,B$ to fit estimated advantage values.}
    		\State {\hspace{1em} Compute policy gradient estimate ${\mathbf{g}}_{\boldsymbol{\theta}_{i}}$ using (\ref{g_estimate}).}
    		\State {\hspace{1em} Compute FIM estimate $\hat{\mathbf{F}}_{\boldsymbol{\theta}_{i}}$ using (\ref{F_estimate})}.
    		\State {\hspace{1em} Compute policy update direction $\Delta_{i}$ using (\ref{delta}).}
    		\State {\hspace{1em} \textbf{for} $j=0,1,\ldots$ \textbf{do}}
    		\State \hspace{2em} Compute: \[{\boldsymbol{\theta}_{i+1}}_{{{}}{{}}{{}}}{=}{\boldsymbol{\theta}}_{{i}}{+}{\mathit{\zeta}}^{j}{\Delta}_{i}\]
    		\State \hspace{2em} \textbf{if} (\ref{L_nonnegative}) and (\ref{KL_constraint_respected}) are satisfied \textbf{then}
    		\State \hspace{3em} Apply policy update at BSs $b=1,\ldots,B$.
    		\State \hspace{3em} break
    		\State \hspace{2em} \textbf{end if}		
    		\State \hspace{1em} \textbf{end for}		
    		\State \hspace{0em}\textbf{end for}
    	\end{algorithmic}
    \end{algorithm}
	
%	\begin{figure}[!t] 
%	\begin{center} 
%		\includegraphics[trim={8cm 3cm 12cm 3cm},clip, height=0.45\textwidth]{fully_decentralized_diagram.pdf}		
%		\caption{}
%		\centering
%		\label{decentralized_diagrams}
%	\end{center}
%	\end{figure} 
\newpage
	\section{Numerical Results}
	To evaluate the performance of the proposed methods, we simulated different cellular networks with the following system parameters:
	\begin{center}
		\captionof{table}{Numerical values of system model parameters} \label{tab1}
		\begin{tabular}{ |c|c| } 
			\hline
			Total bandwidth & \textit{W} = $20$ MHz \\  \hline
			BS maximum transmit power& ${P}_\mathrm{max}$ = $43$ dBm\\ \hline
			Noise PSD & ${N}_{0}$ = $-150$ dBm/Hz \\ \hline
			Noise figure & ${N}_{f}$ = $9$ dB \\ \hline
			Reference distance & $d_0$ = $0.3920$ m \\ \hline
			Cell radius & $1000$ m \\ \hline
		\end{tabular}
	\end{center}
	The parameters utilized for the proposed centralized, partially and fully decentralized deep learning algorithms are chosen identically as follows:
	\begin{center}
		\captionof{table}{Numerical values of deep learning parameters} \label{tab2}
		\begin{tabular}{ |c|c| } 
			\hline
			KL-divergence constraint& ${\delta}$ = 0.01\\ \hline
			Step size & ${\zeta}$ = 0.90 \\ \hline		
			Discount factor & ${\gamma}$ = 0.99 \\ \hline
			Policy network hidden layers &  3\\ \hline		
			Value network hidden layers &  3\\ \hline		
			Neurons per hidden layer&  256\\ \hline			
			Episodes per iteration&  $M$=1000\\ \hline
		\end{tabular}
	\end{center}
	
	Furthermore, the neurons in the hidden layers for both the policy and value networks are chosen to be exponential linear units and have an activation function given by:
	\begin{equation}
	\mathit{\sigma}\left({z}\right){=}\left\{{\begin{array}{c}{\hspace{-0.40em}{z}\hspace{0.33em}\hspace{0.33em}\hspace{0.33em}\hspace{3.70em}{z}{>}{0}}\\{\mathit{}{{e}^{z}{-}{1}}\hspace{0.33em}\hspace{0.33em}\hspace{0.33em}\hspace{1.93em}{z}\leq{0}}\end{array}}\right.
	\end{equation}
	{\color{black}
	The proposed DRL schemes were implemented using Python $3.6$ and TensorFlow $1.14.0$. The machine utilized for training and evaluation of all schemes was equipped with a dual-core $3.1$ GHz Core i$5$ processer, $8$ GB of memory and no discrete GPU.}
\newpage
	\subsection{Training Performance}
		{\color{black}
	We begin by considering training performance for the proposed approaches in a $3$-cell network with the parameters in Table \ref{tab1} and $K=2$ users per cell; additionally, the path loss exponent utilized in this scenario was set to $\alpha=3.76$. Note that we utilize this small network size to fully evaluate training performance and algorithm convergence for all three approaches. The policy and value networks are characterized using the parameters in Table \ref{tab2}. It should be noted that different random initializations of the neural networks may converge to different final policy and value networks; thus, we plot training curves for $10$ different parameter initializations to analyze the convergence behaviour. Furthermore, to ensure a fair comparison, we train the centralized, partially decentralized, and fully decentralized algorithms for $4\times{10}^6$ training steps each; this corresponds to $4\times{10}^6$ and $4/3\times{10}^6$ episodes for the centralized and decentralized approaches respectively. 
	
	Figure $5(a)$ illustrates the evolution of per-episode reward for the centralized approach. Starting from an average of around $40$ Mbps, the algorithm converges to a per-episode reward of $122$ Mbps. Furthermore, we observe that the training curves for the different random initializations are highly consistent in converging to a similar final reward. We observe some variance during the training process; this is to be expected, since the channel realizations (and hence the network sum-rate achieved) during each time slot are generated randomly. The convergence behaviour can be visualized better by plotting an exponentially weighted average of the training performance \footnote{The exponentially weighted average $s[n]$ of a sequence $x[n]$ is calculated using the relation \[s[n]=\left\{ \begin{array}{cc}
			wx[n]+(1-w)s[n-1] & n>1\\
			x[n] & n=1
		\end{array}\right.\]
	\hspace{1.50em}where $w$ represents the smoothing factor. In this paper, we use $w=0.96$.} which is illustrated in Figure $5(b)$. As we can observe, the different random initializations follow closely matching training trajectories in converging to nearly identical final rewards.
	
	In Figures $\ref{pdecent_training}$ and $\ref{fdecent_training}$ respectively, we plot similar training curves for $10$ random initializations of the partially and fully decentralized algorithms. In these cases too, we observe that the average reward improves as training progresses. However, two crucial differences emerge in comparison to the centralized algorithm: first, the decentralized algorithms converge to a lower final reward, and second, they demonstrate notably higher variance across runs. In particular, considering the smoothed training curves, we observe that both the partially and fully decentralized approaches achieve a final reward of around $120$ Mbps. The fully decentralized algorithm also demonstrates the greatest spread of final rewards among the different random initializations and, on average, is outperformed by the partially decentralized algorithm in this setting.
	
\begin{figure}[]
	\vspace{-6.00em}
	\centering
	\begin{subfigure}[]{0.50\textwidth}
		\centering
			\includegraphics[trim={2cm 3cm 0cm 0cm},clip, height=0.79\textwidth]{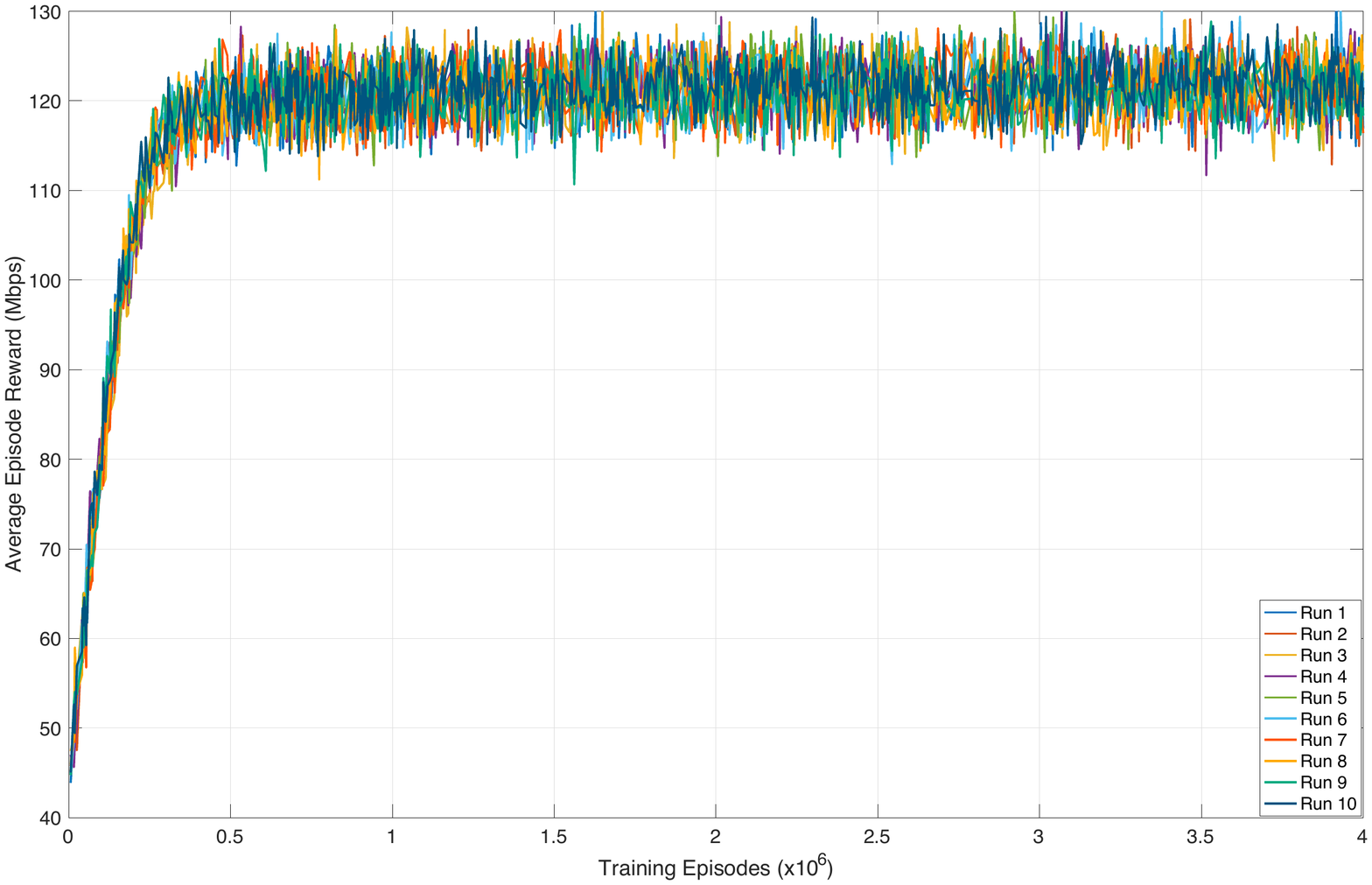}
		\caption{Training convergence.}\label{cent_rough}
	\end{subfigure}\hfill
    \begin{subfigure}[]{0.50\textwidth}
		\centering
		\includegraphics[trim={2cm 3cm 0cm 0cm},clip, height=0.79\textwidth]{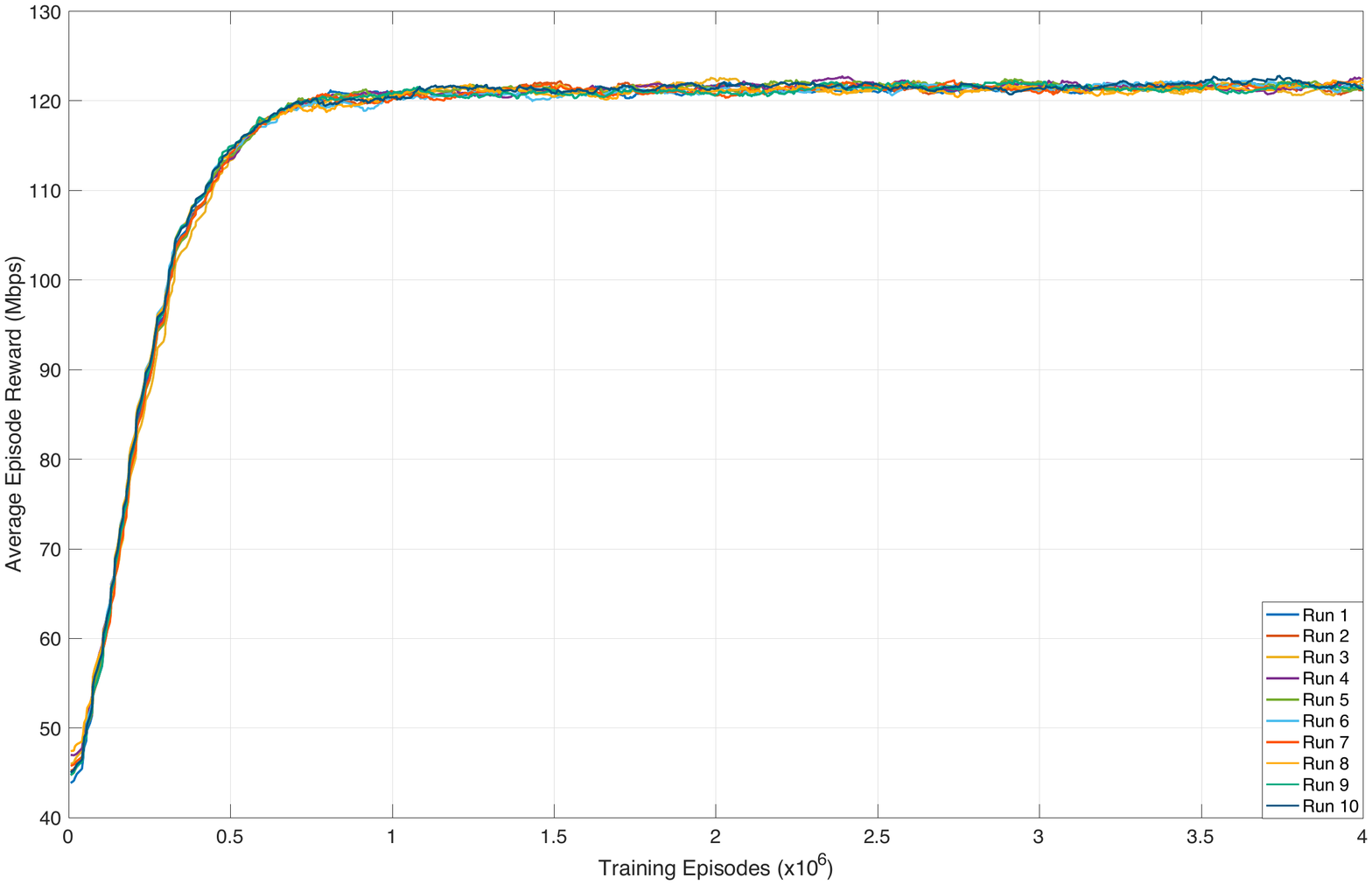}
		\caption{Exponentially weighted average training convergence}\label{cent_smooth}
	\end{subfigure}
	\caption{Centralized training convergence for $10$ random initializations, $B=3$, $K=2$.}
	\centering
	\begin{subfigure}[]{0.50\textwidth}
		\centering
		\includegraphics[trim={2cm 3cm 0cm 0cm},clip, height=0.79\textwidth]{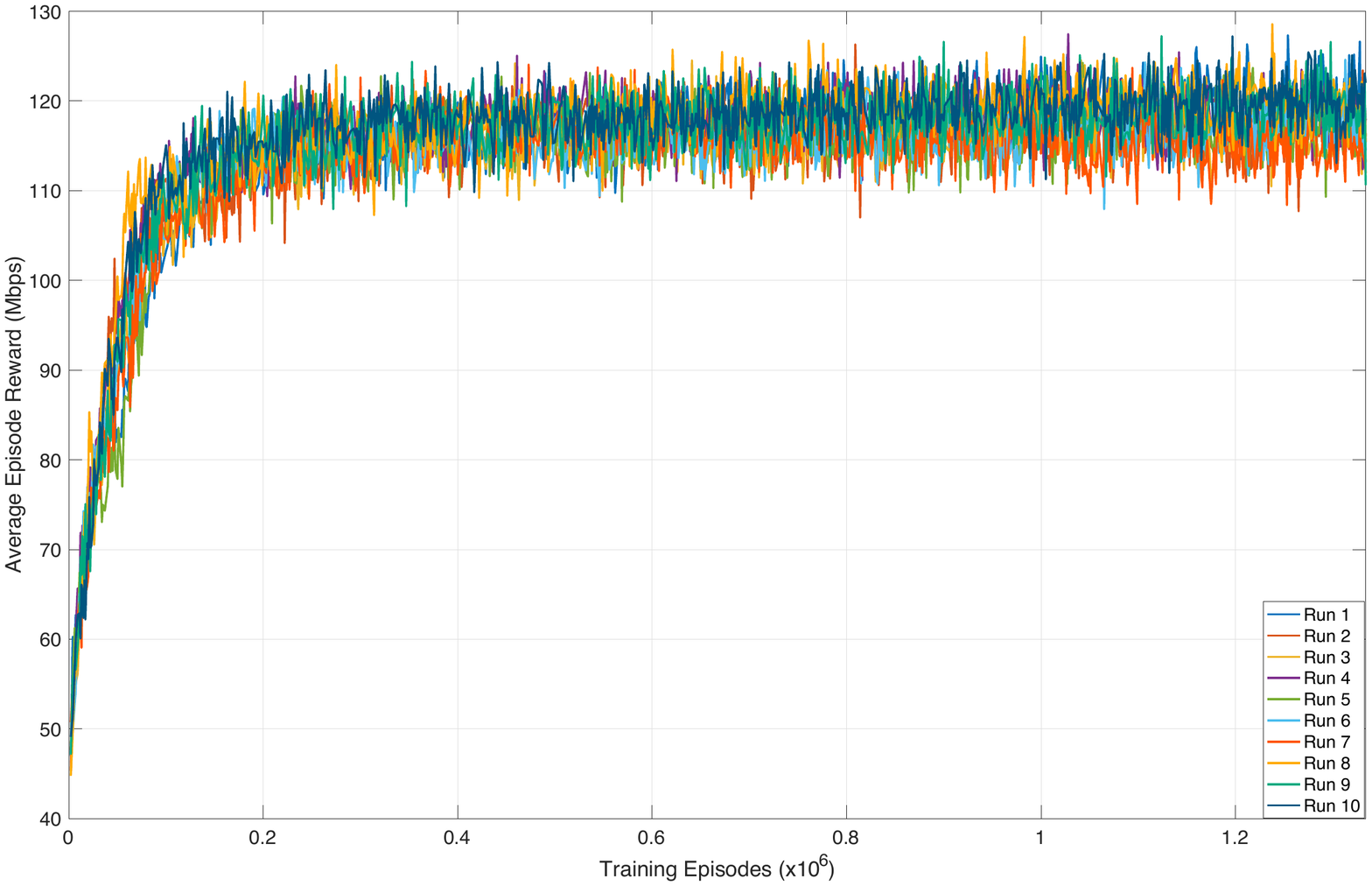}
		\caption{Training convergence.}\label{pdecent_rough}
	\end{subfigure}\begin{subfigure}[]{0.50\textwidth}
		\centering
		\includegraphics[trim={2cm 3cm 0cm 0cm},clip, height=0.79\textwidth]{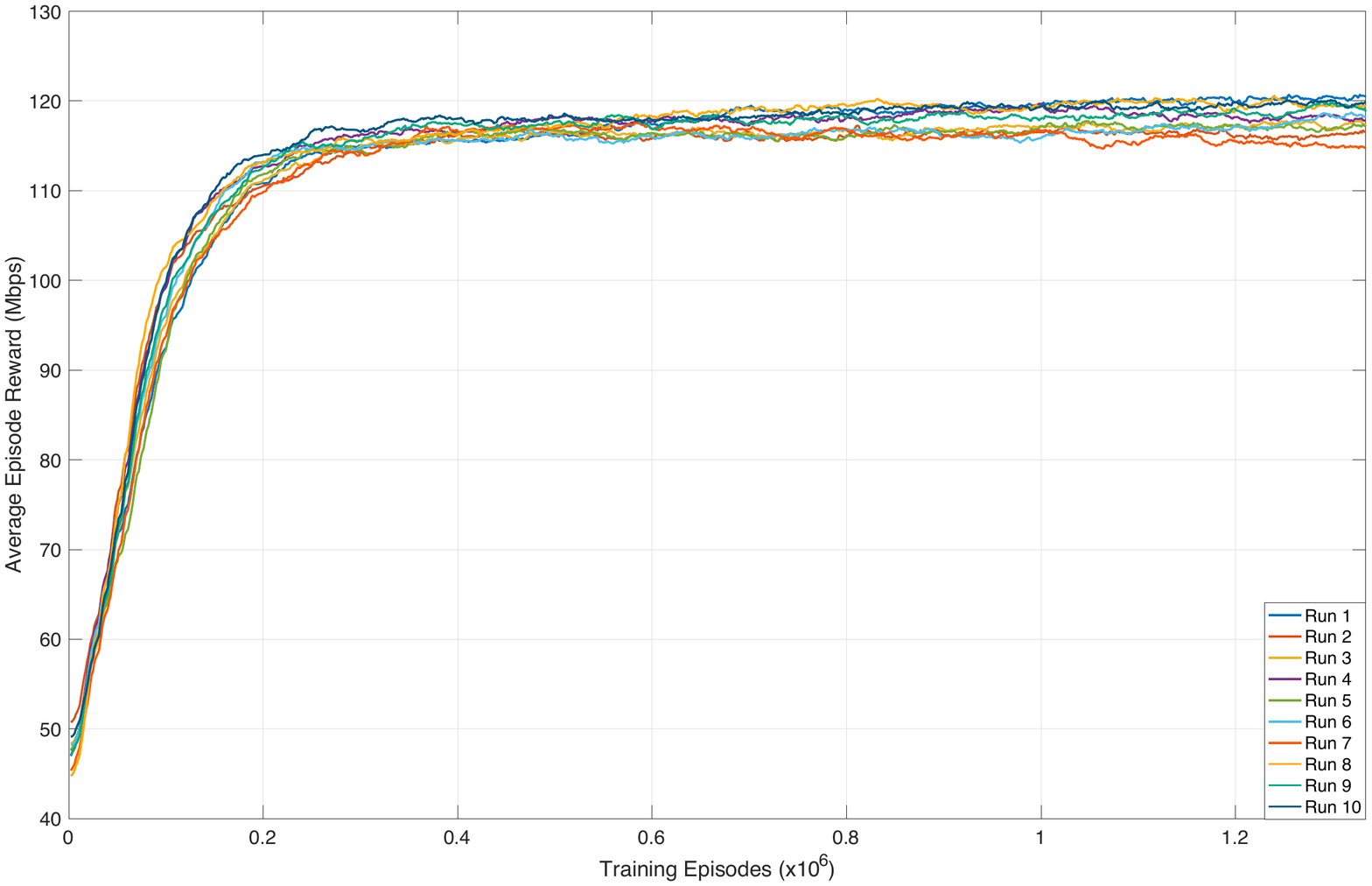}
		\caption{Exponentially weighted average training convergence}\label{pdecent_smooth}
	\end{subfigure}
	\caption{Partially decentralized training convergence for $10$ random initializations, $B=3$, $K=2$.}\label{pdecent_training}\vspace{-0.00em}
	\centering
	\begin{subfigure}[]{0.50\textwidth}
		\centering
		\includegraphics[trim={2cm 3cm 0cm 0cm},clip, height=0.79\textwidth]{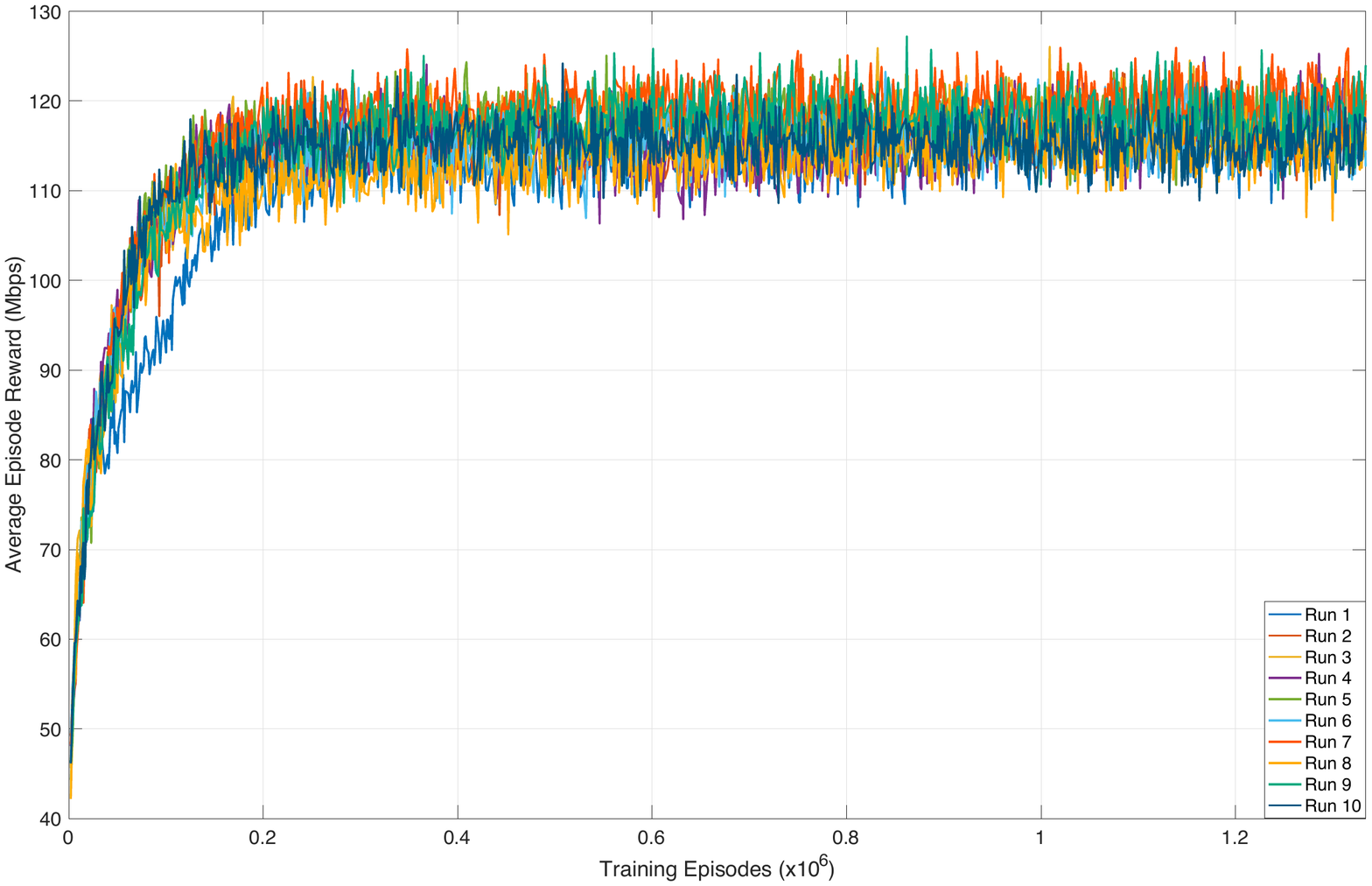}
		\caption{Training convergence.}\label{fdecent_rough}
	\end{subfigure}\begin{subfigure}[]{0.50\textwidth}
		\centering
		\includegraphics[trim={2cm 3cm 0cm 0cm},clip, height=0.79\textwidth]{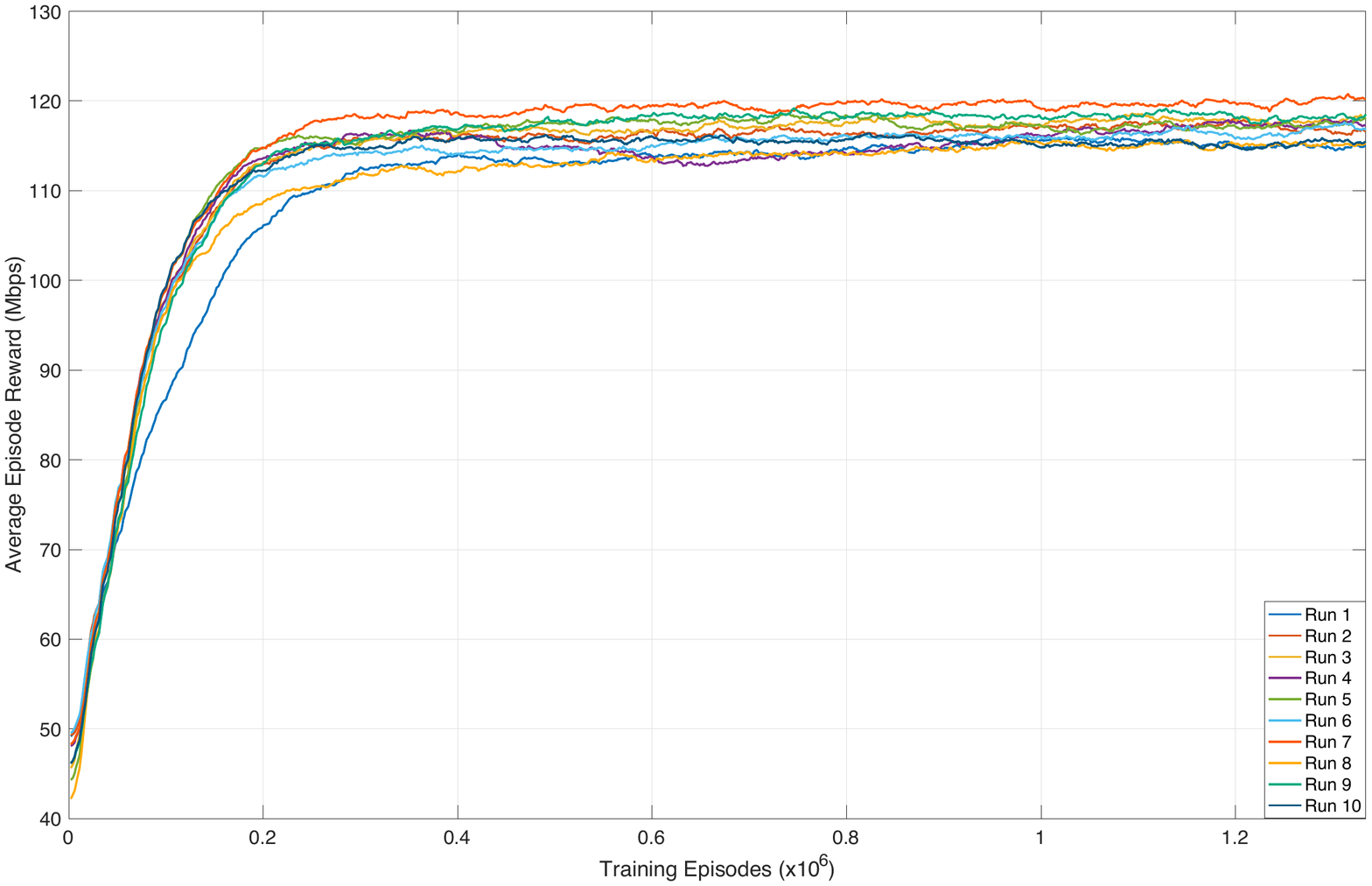}
		\caption{Exponentially weighted average training convergence}\label{fdecent_smooth}
	\end{subfigure}
	\caption{Fully decentralized training convergence for $10$ random  initializations, $B=3$, $K=2$.}\label{fdecent_training}
\end{figure} 

{\color{black}
As a baseline, we also compare the training performance against the well-known A2C (Advantage Actor-Critic) DRL algorithm first proposed in \cite{mnih2016asynchronous}. Like TRPO, A2C utilizes a critic network to help reduce variance in the policy gradient estimates; however, unlike TRPO, the step size is fixed. We plot the training performance of the A2C algorithm for the $3$-cell centralized setting for $10$ random policy initializations with a fixed step size of $7\times{{10}^{-4}}$ and a smoothing factor of $0.96$ in Figure \ref{a2c_training}. The A2C algorithm is unable to converge in this setting and demonstrates poor performance across all initializations. Specifically, the average reward attained fluctuates between approximately $65$ and $82$ Mbps; this is in contrast to the corresponding centralized TRPO method, which achieves an average reward of around $122$ Mbps with a variation of less than $1$ Mbps across all initializations. This can be directly attributed to the fact that taking even small changes in the policy parameters can lead to unexpected changes in terms of the reward function.}
\begin{figure}[!t] 
	\begin{center} 
		\includegraphics[trim={0cm 3cm 0cm 3cm},clip, height=0.55\textwidth]{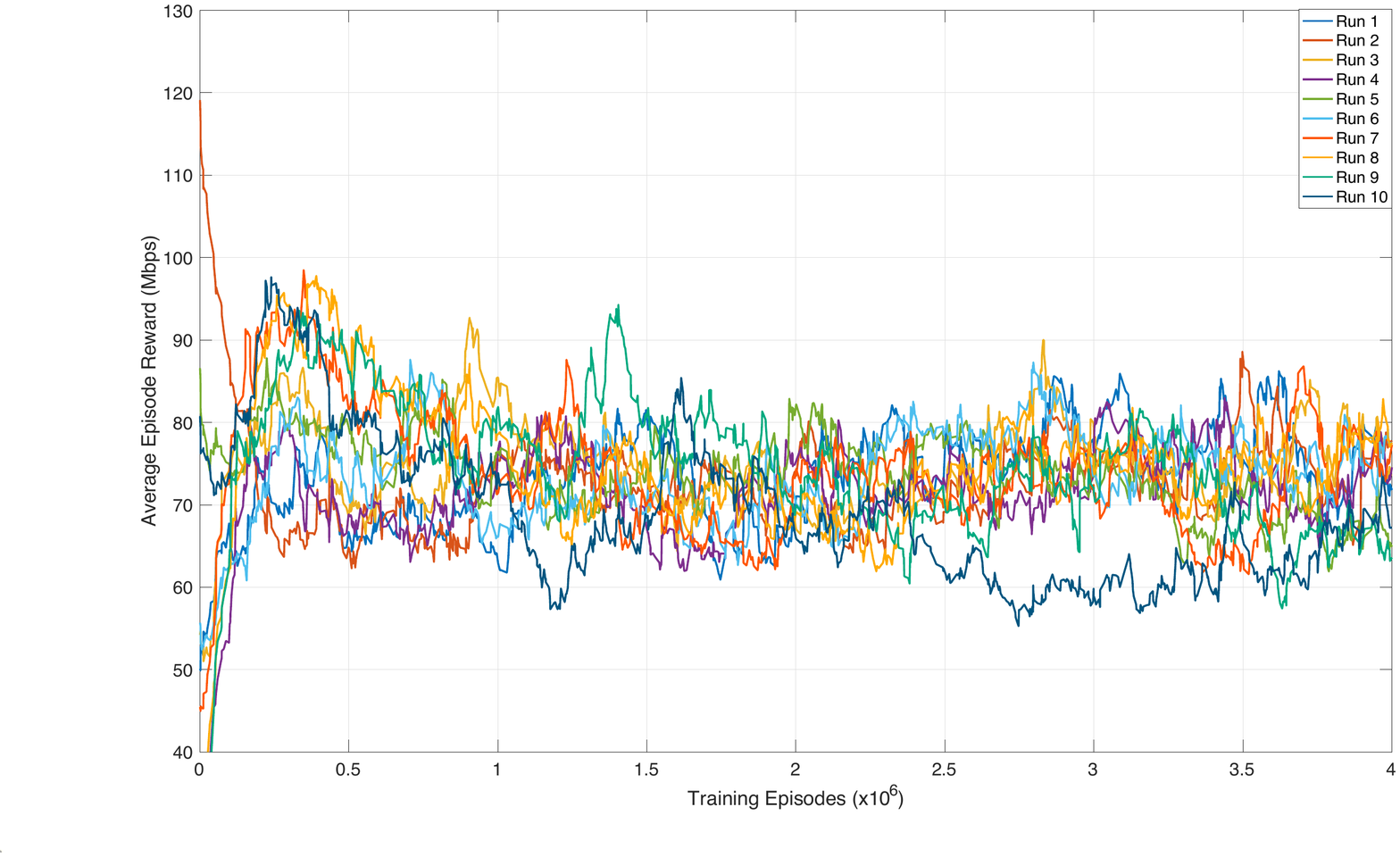}		
		\caption{{\color{black}Exponentially weighted centralized training convergence for A2C algorithm for $10$ random initializations; $B=3$, $K=2$.}}
		\centering
		\label{a2c_training}
	\end{center}
\end{figure} 
	\subsection{Sum-Rate Performance}
    In this section, we compare the performance of the proposed approaches against the following uncoordinated and state-of-the-art coordinated algorithms:
    \begin{enumerate}
    		\item\textit{Maximum Power Allocation}: In this scheme, each BS transmits at the maximum power $P_\mathrm{max}$ to every user within its cell. As the simplest uncoordinated scheme, this scheme necessarily achieves poor performance; nonetheless, it is the least computationally complex and requires no exchange of information between the BSs. Thus, it serves as a useful benchmark to evaluate the performance and complexity of more sophisticated resource management solutions.
    		\item \textit{Random Power Allocation}: For this scheme, the power allocated to each user by its serving base station is chosen uniformly randomly from the interval $[0,P_\mathrm{max}]$. Like the maximum power allocation scheme, this approach is desirable from a computation and information exchange perspective although its performance is worse than that of coordinated schemes.
    		\item \textit{Fractional Programming:} FP is an iterative method based on minorization-maximization developed by Shen and Yu in \cite{shen2018fractional}, and is the highest-performing resource allocation scheme developed thus far in the literature \cite{khan2018optimizing,shen2018fractional,nasir2018deep}. As emphasized in the introduction, FP is guaranteed to converge to a local optimum of the network-sum-rate optimization problem; however, the number of iterations needed to converge is unknown \textit{a priori}.
    		\item \textit{Weighted Minimum Mean-Squared Error}: WMMSE transforms the original sum-rate maximization problem into an equivalent optimization problem of minimum mean-squared error minimization. Like FP, WMMSE is guaranteed to converge in a monotonically nondecreasing fashion to a local optimum of the problem. Specifically, we compare our proposed methods against the SISO variant of the WMMSE algorithm as utilized in \cite{nasir2018deep,sun2018learning,shen2018fractional,Meng2019}.
    \end{enumerate}
    Note that we do not compare against the decentralized approaches in \cite{de2015best,fritzsche2013distributed,park2010distributed,menache2011network} as we consider the state-of-the-art FP and WMMSE algorithms as the performance benchmarks.
    
    We begin by considering the performance of the different algorithms for the aforementioned $B=3$ cells, $K=2$ users per cell setting considered for the training curves. Figure \ref{sum_rate_convergence_3cell} plots the convergence of the network sum-rate for a single random initialization of channel values according to the given system model. As expected, the uncoordinated max-power and random-power allocation schemes achieve the worst performance; on the other hand, both FP and WMMSE converge in a monotonically nondecreasing fashion to local optima of the network sum-rate maximization problem as expected. In comparison, both the coordinated and uncoordinated schemes are outperformed by the proposed deep learning approaches. As with the training results, the centralized approach outperforms the partially decentralized approach which, in turn, achieves a small performance advantage over the fully decentralized scheme. Also, the coordinated FP and WMMSE algorithms require multiple iterations to reach the local optimum of the problem; this is in contrast to the proposed DRL approaches, in which the state information can be directly propagated through the policy network to obtain the desired solution to the optimization problem.
	
	\begin{figure}[!t] 
		\begin{center} 
			\includegraphics[trim={0cm 3cm 0cm 3cm},clip, height=0.55\textwidth]{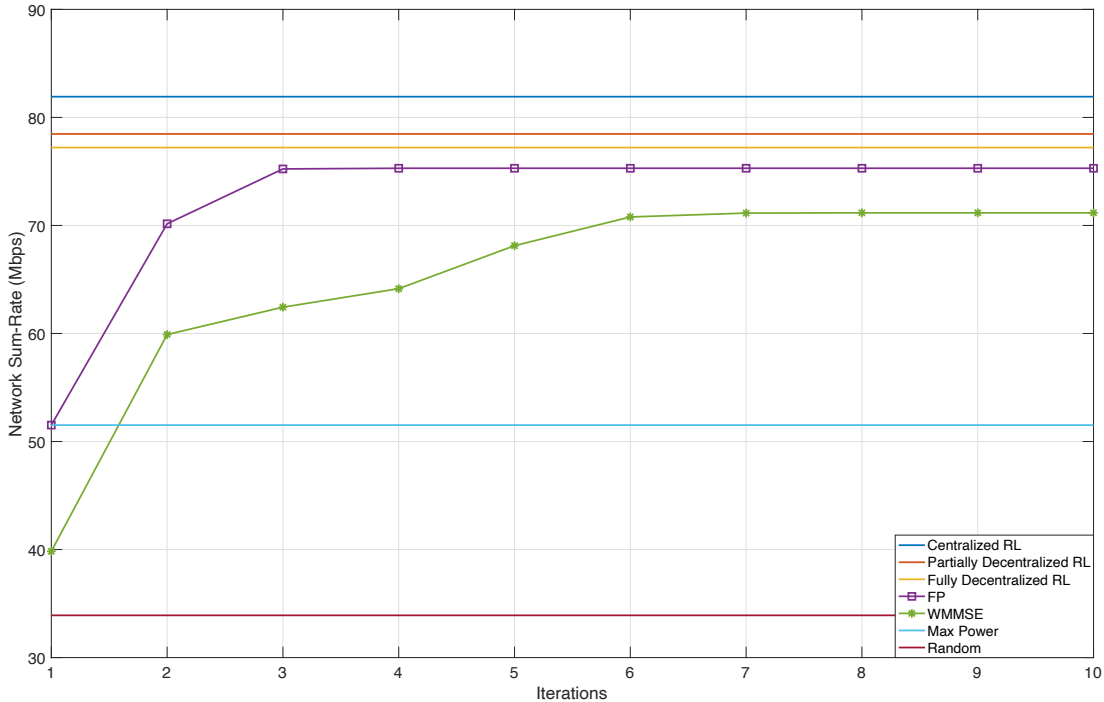}
			\caption{Sum-rate convergence for a single random channel realization, $B=3$, $K=2$.}
			\centering
			\label{sum_rate_convergence_3cell}
		\end{center}
	\end{figure} 
	
	The results in Figure $\ref{sum_rate_convergence_3cell}$ are for a single time slot with user locations and channels generated randomly according to the distributions described in the system model. To compare fairly with the stated benchmarks, in Figure $\ref{averaged_sumrates}$ we consider the average network sum-rate achieved across $1000$ independent random channel realizations. These results display a similar trend: the centralized scheme achieves the highest average network spectral efficiency, followed by the partially and fully decentralized schemes respectively. The model-based optimization algorithms are outperformed by the DRL schemes while the uncoordinated schemes once again achieve the worst performance.
	\begin{figure}[!h] 
		\begin{center} 
			\includegraphics[trim={0cm 6cm 0cm 7cm},clip, height=0.55\textwidth]{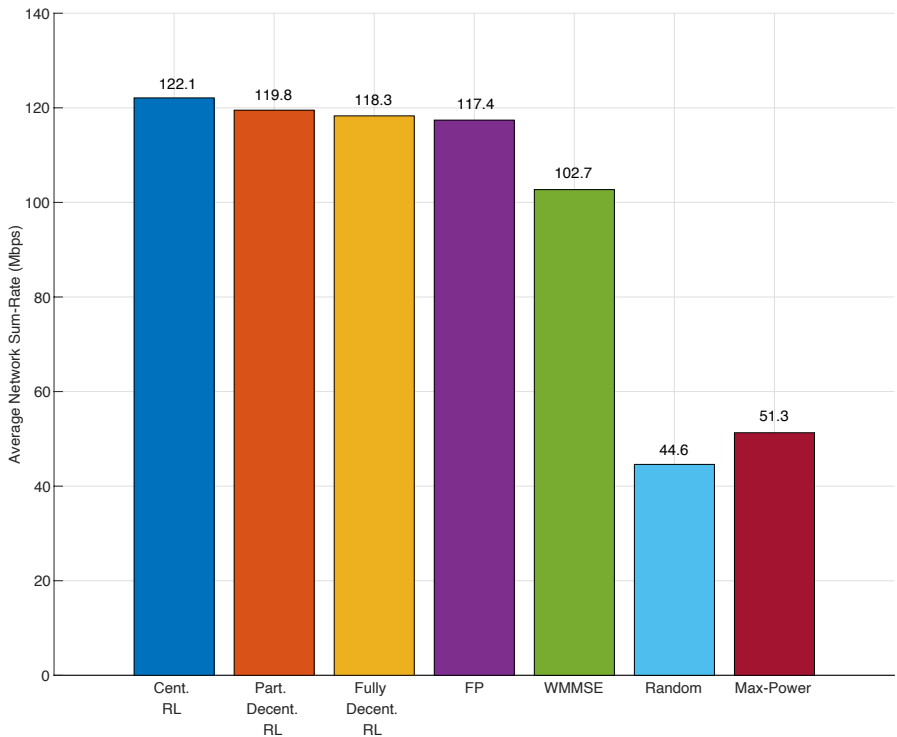}
			\caption{Averaged network sum-rate for $1000$ channel realizations, $B=3$, $K=2$.}
			\centering
			\centering
			\label{averaged_sumrates}
		\end{center}
	\end{figure} 
We also remark that even the fully decentralized approach outperforms the state-of-the-art FP and WMMSE algorithms, despite the fact that it uses only a fraction of the downlink CSI available to these coordinated schemes. This result is all the more remarkable when we consider that none of the proposed approaches are trained for particular channel realizations; instead, during training, the user locations and channel values are generated randomly from the stated distributions in the channel model. In other words, these results indicate that the DRL approaches achieve good generalization, as they are able to perform well on channel realizations that are drawn from the same distribution as, but \textit{not identical to}, the training episodes \cite{bishop2006pattern}.
	
As stated earlier, the chief drawback of the centralized scheme is that the training times become impractically large compared to the decentralized schemes as the cellular network size grows. Furthermore, the information exchange requirements are identical to the FP and WMMSE algorithms and as these become burdensome for larger wireless networks. Accordingly, we proceed by testing the performance of the decentralized approaches in a much larger $7$-cell hexagonal network with wraparound; the network parameters are chosen to be identical to the $3$-cell setting, with the exception of the pathloss exponent $\alpha$ which is set to $4.00$. Figure $\ref{sum_rate_convergence_7cell}$ demonstrates the convergence of network sum-rate for a single random realization of channels for this setting; similar to the 3-cell network, the partially and fully decentralized approaches once again achieve higher network spectral efficiency than the benchmark schemes.
	
	\begin{figure}[!t] 
		\begin{center} 
			\includegraphics[trim={0cm 3cm 0cm 3cm},clip, height=0.55\textwidth]{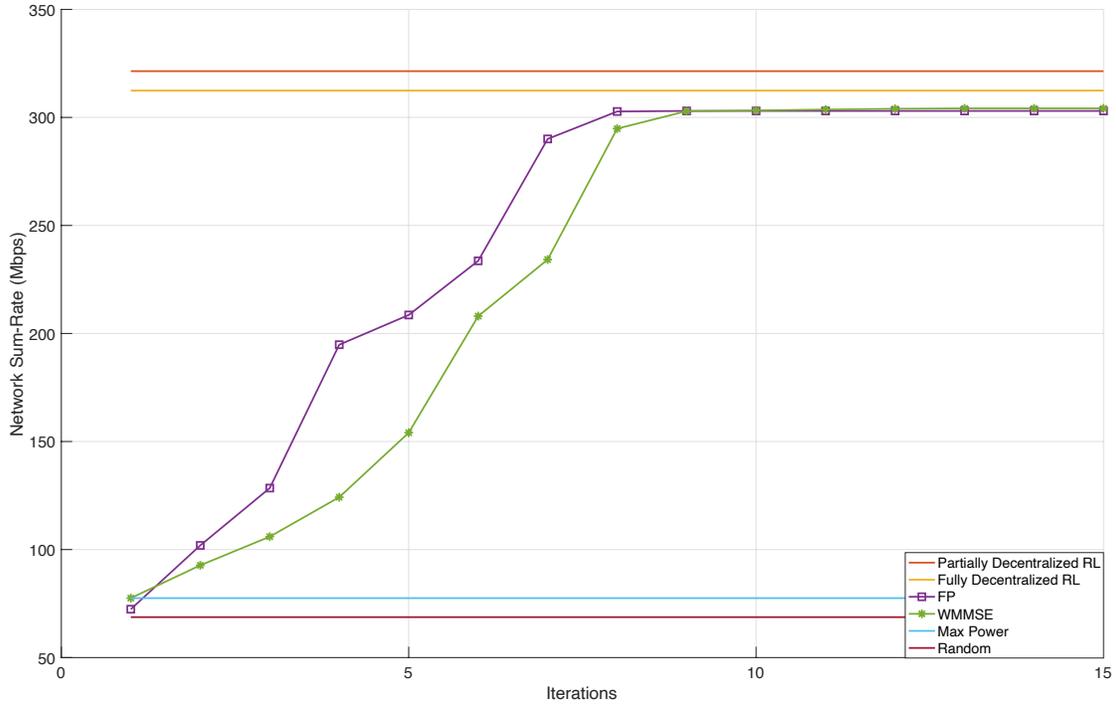}
			\caption{Sum-rate convergence for a single random channel realization, $B=7$, $K=8$.}
			\centering
			\centering
			\label{sum_rate_convergence_7cell}
		\end{center}
	\end{figure} 
	We also consider the average sum-rate achieved across a large number of independent channel realizations in Figure $\ref{averaged_sumrates_7cell}$, similar to Figure $\ref{averaged_sumrates}$. These results display similar trends to the $3$-cell setting, with the partially decentralized approach once again achieving the highest network spectral efficiency and achieving 7\% higher network sum-rate than FP. The fully decentralized approach is once again outperformed by the partially decentralized approach, but still manages to achieve higher objective function values than the FP, WMMSE, random and max-power schemes.
	\begin{figure}[!t] 
		\begin{center} 
			\includegraphics[trim={0cm 6cm 0cm 7cm},clip, height=0.55\textwidth]{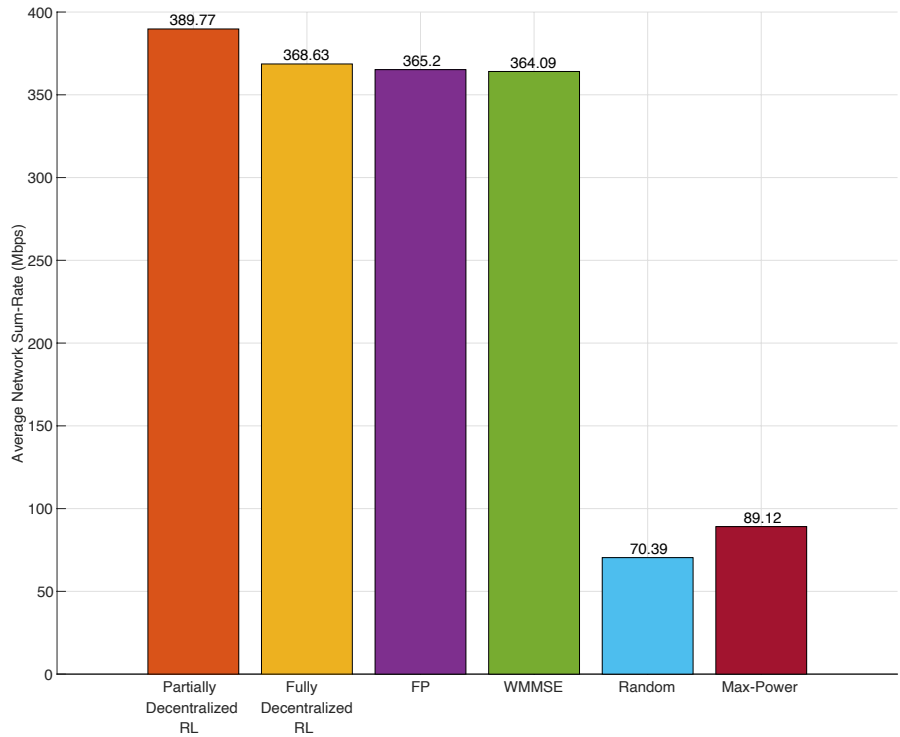}
			\caption{Averaged network sum-rate for $1000$ channel realizations, $B=7$, $K=8$.}
			\centering
			\centering
			\label{averaged_sumrates_7cell}
		\end{center}
	\end{figure}

%To further assess the performance, we also simulate an even larger network size consisting of $16$ cells, with $K=8$ users per cell. For this setting, we utilize a different topology, consisting of square cells with a width of $2000$m each arranged in a $4\times4$ grid. The results of the average network sum-rate achieved across multiple independent channel realizations are illustrated in Figure ([REFERENCE]). Interestingly, we observe that the performance margin of both the partially and fully decentralized approaches over FP and WMMSE algorithms increases, demonstrating further the advantages of our approach.
To test the robustness of the proposed RL approaches we also plot the average sum-rate achieved across multiple time slots for $P_{max}$ values ranging between $20$ and $50$ dBm for the $3$-cell setting in Figure $\ref{TX_powers}$. For this test, we utilize the policy networks that have been previously trained for a transmit power of $43\hspace{0.33em}\mathrm{dBm}$ by simply scaling the output of the policy networks to match the transmit power constraint; this tests the performance of the DRL approaches when the reward function is changed from that utilized during training. As we can see, all three DRL methods demonstrate excellent performance across the entire range, closely matching the FP and WMMSE algorithms for the lower transmit values. At higher transmit power levels, the proposed DRL approaches demonstrate superior performance to the WMMSE algorithm while continuing to match the FP algorithm.
	\begin{figure}[!t] 
	\begin{center} 
		\includegraphics[trim={0cm 3cm 0cm 3cm},clip, height=0.55\textwidth]{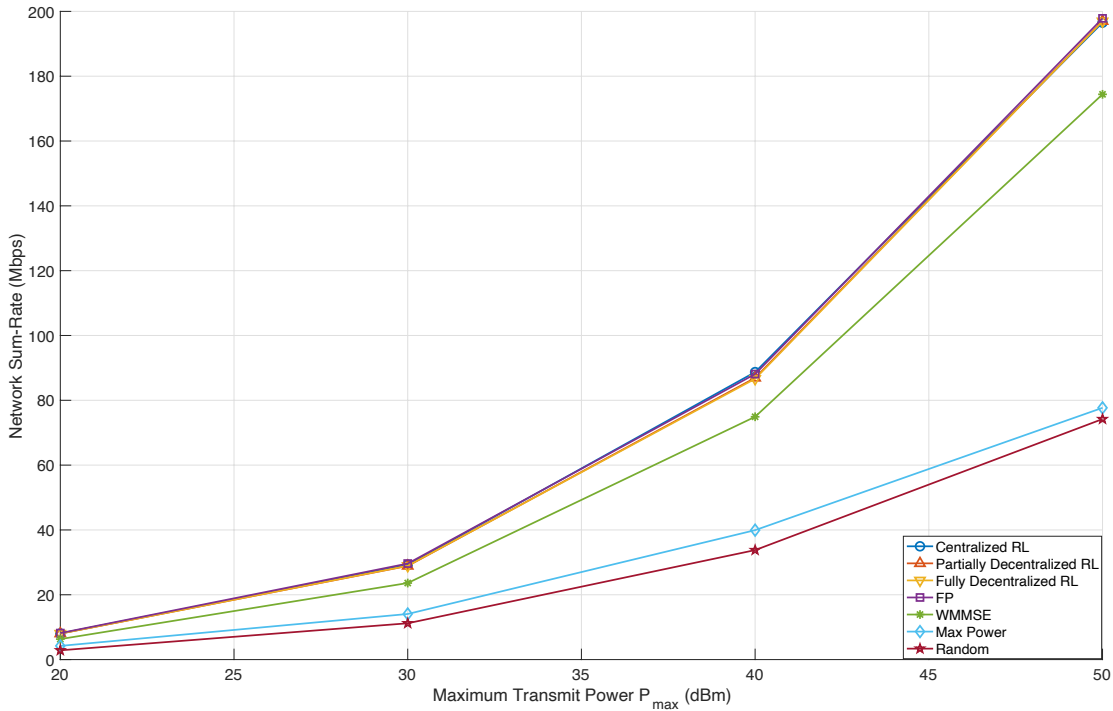}
		\caption{Averaged network sum-rate for different transmit powers, $B=3$, $K=2$.}
		\centering
		\centering
		\label{TX_powers}
	\end{center}
\end{figure}

For completeness, we also consider a slight variation of the original network sum-rate maximization problem with a sum-power constraint of $43$ dBm instead of a per-user power constraint for the $3$-cell setting. This power constraint is enforced for the RL algorithms by simply rescaling the power output during training to ensure that the per-BS sum power does not exceed $P_{max}$. As the results in Figure $\ref{sum_power}$ show, the fully centralized and partially decentralized RL algorithms still outperform the FP and WMMSE algorithms, as well as the uncoordinated approaches.
\begin{figure}[!t] 
	\begin{center} 
		\includegraphics[trim={0cm 6cm 0cm 7cm},clip, height=0.55\textwidth]{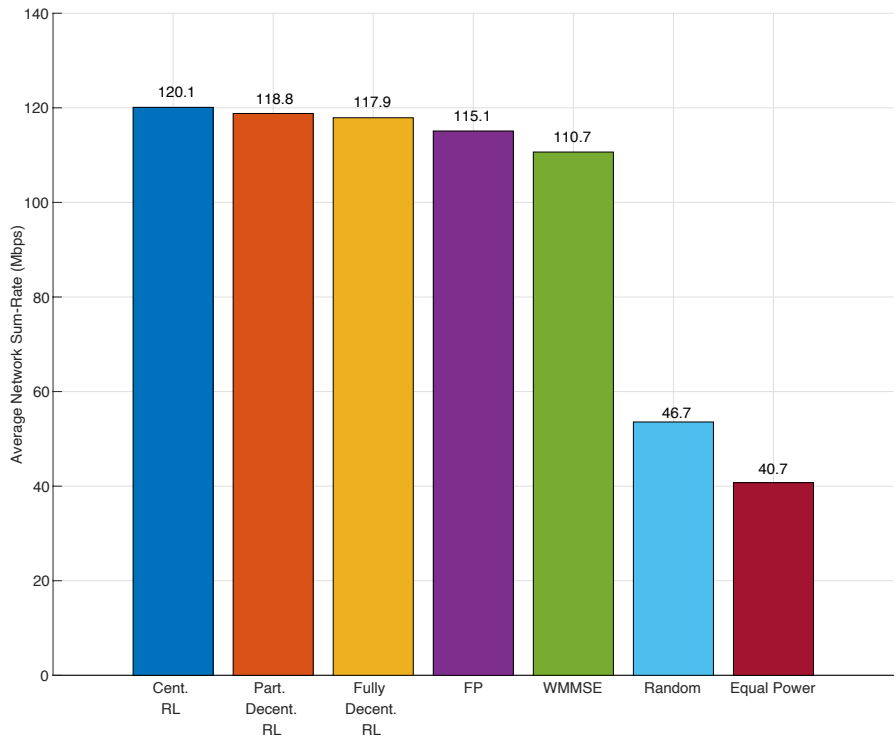}
		\caption{Averaged network sum-rate for sum-power constraint, $B=3$, $K=2$.}
		\centering
		\centering
		\label{sum_power}
	\end{center}
\end{figure}

\subsection{Channel State Information Exchange}
In comparing the performance of different schemes, it is important to consider the amount of information exchange between BSs necessary in order to achieve the resulting performance. As mentioned earlier, the FP and WMMSE algorithms require all BSs to send their complete downlink CSI to a central cloud processor which then computes and forwards  in order to compute the power allocation variables. For a network consisting of $B$ cells and $K$ users per cell, this corresponds to the central processor receiving ${\mathcal{O}}\left({{KB}^{2}}\right)$ scalars from the cooperating BSs. The centralized single-agent approach also requires identical CSI exchange between the BSs. 
	
For the partially decentralized approach, the state information for each BS comprises the downlink CSI from itself to all the users in the network as well as the power allocation decisions of the previous BSs. This corresponds to an information exchange requirement of ${\mathcal{O}}\left({KB}\right)$. On the other hand, for the fully decentralized approach, the state information comprises only the downlink channels from itself to the users in the network. Accordingly, there is no information exchange between the BSs of the network. This is similar to the uncoordinated max-power and random-power schemes. These results are summarized in Table \ref{tab3}, along with the necessary CSI per BS.
	
\begin{table}[h!]
		\captionof{table}{Comparison of CSI and information exchange requirements} \label{tab3}
		\begin{tabular}{|l|l|l|}
			%\rowcolor[HTML]{EFEFEF} 
			\hline\textrm{\begin{tabular}[c]{@{}l@{}}Resource Allocation Scheme\end{tabular}} & \textrm{\begin{tabular}[c]{@{}l@{}}Information exchange\end{tabular}} & \textrm{\begin{tabular}[c]{@{}l@{}}Per-BS CSI needed\end{tabular}} \\ \hline \hline
			Centralized RL                      & ${\mathcal{O}}\left({{KB}^{2}}\right)$                     &${\mathcal{O}}\left({{KB}^{2}}\right)$                          \\ \hline
			Partially Decentralized RL          & ${\mathcal{O}}\left({{KB}}\right)$                     & ${\mathcal{O}}\left({{KB}}\right)$                          \\ \hline
			Fully Decentralized RL              & $0$                     &   ${\mathcal{O}}\left({{KB}}\right)$                         \\ \hline
			FP                                  & ${\mathcal{O}}\left({{KB}^{2}}\right)$                     & ${\mathcal{O}}\left({{KB}^{2}}\right)$                         \\ \hline
			WMMSE                               & ${\mathcal{O}}\left({{KB}^{2}}\right)$                     & ${\mathcal{O}}\left({{KB}^{2}}\right)$                          \\ \hline
			Max-power                           & $0$                     & $0$                          \\ \hline
			Random                              & $0$                     & $0$      \\ \hline                   
		\end{tabular}
	\centering
	\end{table}
	\subsection{Execution Times}
	An essential aspect of comparison between different resource allocation algorithms is to consider the time and complexity necessary to calculate a solution. A key benefit of utilizing deep reinforcement learning is that propagating the current state through the policy network to find the action is typically quite fast; thus, when the agents have been trained, power allocation solutions can be found in a fraction of the time required by conventional optimization techniques. 
	
	Table \ref{tab4} illustrates the execution times of the trained models for our implementations of the different schemes. As we can observe, the deep-RL algorithms require an execution time that is over two orders of magnitude lower to calculate a power allocation strategy as compared to the WMMSE and FP algorithms. All times have been measured for the $3$-cell setting; although since we use identical neural network sizes throughout, these results are equally representative for the $7$-cell network.
	
	It should be noted that the partially decentralized approach requires the longest execution time among the DRL approaches due to the sequential sharing of power allocation information. The state information for each BS needs to be propagated through its policy network to produce its power allocation strategy, and then forwarded to the next BS, which repeats this process, and so on; it follows that the last BS would have to wait until all other BSs have computed their power allocation strategies. Accordingly, the execution time for this approach will be $B$ times longer than the fully decentralized approach, which can be executed in parallel at each BS. It should be noted, however, that the latency introduced in this regard would still be much lower than that required by the FP and WMMSE algorithms; for example, projecting from the execution times in Table \ref{tab3}, the execution time would still be around an order of magnitude lower than the FP and WMMSE algorithms.
	\begin{table}[]\
		\captionof{table}{Average execution times of different resource allocation schemes} \label{tab4}
		\begin{tabular}{|l|l|}
			\hline
%			\rowcolor[HTML]{EFEFEF} 
			\textrm{\begin{tabular}[c]{@{}l@{}}Resource Allocation Scheme\end{tabular}} & \textrm{\begin{tabular}[c]{@{}l@{}}Average Execution Time (seconds)\end{tabular}} \\ \hline \hline
			Centralized RL & ${6}{.}{5}\times{10}^{{-}{4}}$ \\ \hline
			Partially Decentralized RL & ${1}{.}{99}\times{10}^{{-}{3}}$ \\ \hline
			Fully Decentralized RL & ${6}{.}{5}\times{10}^{{-}{4}}$ \\ \hline
			FP & ${6}{.}{4}\times{10}^{{-}{1}}$ \\ \hline
			WMMSE & ${6}{.}{0}\times{10}^{{-1}}$ \\ \hline
			Max-power & $0$ \\ \hline
			Random & ${1}{.}{75}\times{10}^{{-}{4}}$ \\ \hline
		\end{tabular}
	\centering
	\end{table}

	\section{Conclusions}
	{\color{black}
	In this paper, we employed a deep-reinforcement learning approach to directly solve the continuous-valued, non-convex and NP-hard downlink sum-rate optimization problem. We proposed multiple variants: a fully centralized single-agent approach as well as partially and fully decentralized multi-agent approaches. The centralized and partially decentralized TRPO approaches achieve higher network sum-rate than the state-of-the-art FP, WMMSE and A2C algorithms while, once trained, all three approaches are capable of finding solutions in a fraction of the time needed by the conventional optimization methods. Of greater consequence, the framework of trust region policy optimization allows us to design decentralized schemes enabling varying degrees of information exchange between base stations while overcoming the `curse-of-dimensionality' issues associated with centralized reinforcement learning.}
	
	This work represents a preliminary step in investigating centralized and decentralized reinforcement learning algorithms for solving wireless resource management problems, and many fruitful directions are possible for future research. In particular, reducing the number of samples and computation necessary to achieve effective performance is possibly the most important open problem in reinforcement learning algorithms. Our work is, to the best of our knowledge, the first to implement information-sharing between DRL agents to solve optimization problems. While the exchange of power information is beneficial in helping increase network spectral efficiency as compared to the fully decentralized scheme, it is possible that feature engineering may be able to improve performance further. Finally, we note that the extension to multiple-input multiple-output (MIMO) wireless networks remains challenging for machine learning algorithms due to input representation and prohibitively large input dimensionality; distributed DRL algorithms such as those proposed in this work may help overcome the latter obstacle.

%	{\color{black}
%	\begin{spacing}{2.0}
		\bibliographystyle{IEEEtran}
		\bibliography{IEEEabrv,biblio}	
		%\bibliography{strings,refs}
%	\end{spacing}}
	
\end{document}